\documentclass[twocolumn,5p]{elsarticle} 

\usepackage{xspace}
\usepackage{tcolorbox}
\usepackage{svg}
\usepackage{amsmath}
\usepackage{tabularx}
\usepackage{float}
\usepackage{url}
\usepackage{alltt}
\usepackage[hidelinks]{hyperref}
\usepackage{cleveref}
\usepackage{subcaption}
\usepackage{listings}
\usepackage{xcolor}
\usepackage[binary-units]{siunitx}
\usepackage{algorithm}
\usepackage{algpseudocode}
\usepackage{stmaryrd}
\usepackage{balance}
\usepackage{enumitem}
\usepackage{colortbl}
\usepackage{bbding}
\usepackage[switch]{lineno}
\usepackage[normalem]{ulem}
\usepackage{fnpct}
\usepackage{multirow}
\usepackage{afterpage}
\usepackage{caption}
\usepackage{longtable}

\DeclareGraphicsExtensions{.pdf, .png, .jpg}
\graphicspath{{pictures/}}

\newcommand\etal[0]{\emph{et al.}\xspace}
\newcommand{\ie}{\emph{i.e.},\xspace}
\newcommand{\eg}{\emph{e.g.},\xspace}

\newcommand{\etc}{\emph{etc.}\xspace}
\newcommand{\wrt}{w.r.t.\xspace}
\newcommand\kb[1]{\SI{#1}{\kilo\byte}\xspace}
\newcommand\mb[1]{\SI{#1}{\mega\byte}\xspace}
\newcommand\gb[1]{\SI{#1}{\giga\byte}\xspace}
\newcommand\pc[1]{\SI{#1}{\percent}\xspace}

\newcommand{\feature}[1]{\textsf{#1}\xspace}
\newcommand{\performance}[1]{\underline{#1}\xspace}
\newcommand{\soft}[1]{\emph{#1}\xspace}
\newcommand{\video}[1]{video~\#\num{#1}\xspace}
\newcommand{\resp}{resp.\xspace}
\newcommand{\bluerl}[1]{\textcolor{blue}{\url{#1}}}

\newcommand{\nbsystems}{\num{8}\xspace}
\newcommand{\nbpapers}{\num{65}\xspace}
\newcommand{\nbsolutionpapers}{\num{15}\xspace}

\newcommand{\rqcorr}{To what extent are the performance distributions of configurable systems changing with input data?}
\newcommand{\rqfeature}{To what extent the effects of configuration options are consistent with input data?}

\newcommand{\rqimpact}{How much performance are lost when reusing a configuration across inputs?}
\newcommand{\rqgroup}{What is the benefit of grouping the inputs?}
\newcommand{\pqone}{Is there a software system processing input data in the study?} 
\newcommand{\pqtwo}{Does the experimental protocol include several inputs?} 
\newcommand{\pqthree}{Is the problem of input sensitivity mentioned in the paper?}
\newcommand{\pqfour}{Does the paper propose a solution to generalize the performance model across inputs?}

 \newcommand{\scriptsizeb}[1]{\scalebox{0.68}{{\normalsize #1}}}

\begin{document}
\begin{frontmatter}

%\title{Performance of Configurable Systems:\\ An Empirical Study on Input Sensitivity}
% \title{Exploring the Input Sensitivity of Configurable Systems \\ An Empirical Study} %% performance not appearing
% il me semble qu'on doit changer le titre pour des raisons de double blind
% à changer donc, mais on pourra reprendre l'ancien titre si c'est accepté
%\title{The Interaction between Inputs and Configurations fed to Software Systems: an Empirical Study} 
%\title{Predicting Performances of Configurable Systems: the Issue of Input Sensitivity} 
%\title{On the Input Sensitivity of Configurable Systems}
% Deep variability: an empirical study on the sensitivity of configuration options to input data
%\title{Sensitivity of Inputs to the Performance of Configurable Systems \\ An Empirical Study}
\title{Input Sensitivity on the Performance of Configurable Systems: An Empirical Study} %% performance not appearing
%\title{The Interplay of Compile-time and Run-time Options for Performance Prediction and Optimization}

\author[firstaddress]{Luc Lesoil}

\author[firstaddress]{Mathieu Acher}
%\ead{mathieu.acher@irisa.fr}

\author[firstaddress]{Arnaud Blouin}

\author[firstaddress]{Jean-Marc Jézéquel}
%\ead{Jean-Marc.Jezequel@irisa.fr}

\address[firstaddress]{Univ Rennes, Inria, INSA Rennes, CNRS, IRISA, France}

\begin{abstract}
Widely used software systems such as video encoders are by necessity highly configurable, with hundreds or even thousands of options to choose from. 
Their users often have a hard time finding suitable values for these options (\ie finding a proper configuration of the software system) to meet their goals for the tasks at hand, \eg compress a video down to a certain size. 
One dimension of the problem is of course that performance depends on the input data: \eg a video as input to an encoder like \soft{x264} or a file  fed to a tool like \soft{xz}.
To achieve good performance, users should therefore take into account both dimensions of (1) software variability and (2) input data.
This paper details a large study over \nbsystems configurable systems that quantifies the existing interactions between input data and configurations of software systems. 
The results exhibit that (1) \textbf{inputs fed to software systems can interact with their configuration options in non-monotonous ways}, significantly impacting their performance properties (2) input sensitivity can challenge our knowledge of software variability and question the relevance of performance predictive models for a field deployment. 
Given the results of our study, we call researchers to address the problem of input sensitivity when tuning, predicting, understanding, and benchmarking configurable systems.
%which makes the configuration process even more challenging. 
%In this problem-statement paper, . %, 
% but (3) can be used to tune software systems for their input data. 
%\wrt its input data.
%\fixme{remplacer par le \% de RQ3?}
% we can gain in average \pc{25} performance.
% It also has practical implications for researchers training performance models and question their generalization across inputs. 
%input{sections/00-Abstract}
\end{abstract}


\end{frontmatter}

% \linenumbers

% \vspace*{-0.1cm}
\section{Introduction}
\label{sec:intro}

%%%%%% CONTEXT
Widely used software systems are by necessity highly configurable, with hundreds or even thousands of options to choose from.
According to Svahnberg \etal~\cite{svahnberg2005}, software variability is the "ability of a software system or artifact to be efficiently extended, changed, customized or configured for use in a particular context".
%End-users, developers, or scientists can combine configuration options to obtain the functionalities and performance they want.
% Examples of such sw and how they are configured. To shorten
For example, a tool like \soft{x264} offers \num{118} run-time options such as \mbox{-}\mbox{-}\feature{ref}, \mbox{-}\mbox{-}\feature{no-mbtree} or \mbox{-}\mbox{-}\feature{no-cabac} for encoding a video. 
The same applies to \soft{Linux} kernels or compiler such as \soft{gcc}: 
they all provide configuration options through compilation options, feature toggles or command-line parameters. 
%%%%%% CHALLENGE
% Say why so complex cannot master configurations manually
Software engineers often have a hard time finding suitable values for those options (\ie finding a proper configuration of the software system) to meet their goals for the tasks at hand, \eg compile a program into a high-performance binary or compress a video down to a certain \performance{size} while keeping its perceived \performance{quality}. 

Since the number of possible configurations grows exponentially with the number of options, even experts may end up recommending sub-optimal configurations for such complex software \cite{DBLP:journals/corr/JamshidiC16}. 

\begin{figure}[htb]
    \centering
    % \vspace*{-0.05cm}
    \includegraphics[width=\columnwidth]{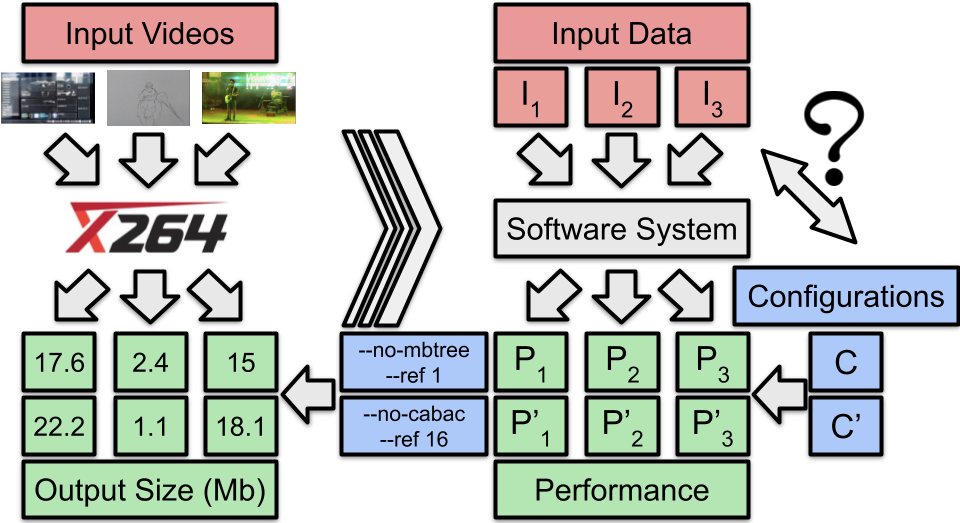}
    %\small
    %\textbf{}
    % \vspace*{-0.25cm}
    \caption{This paper explores and quantifies how inputs fed to software systems impact performance of configurations.}
    \label{fig:intro}
    % See here for modifications:
    % https://docs.google.com/presentation/d/1Lek6qAWCRu1wnQMMaisvPBcSX_I-iChiRZFnAjtVqXA/edit?usp=sharing
\end{figure}

%\fixme{However, there exist cases where inputs (\eg files fed to an archiver like \soft{xz} or SAT formulae provided as input to a solver like \soft{lingeling}) can also impact the performance of a configurable system \cite{xu2008}.}
However, there exists cases where inputs (\eg files fed to an archiver like \soft{xz} or SAT formulae provided as input to a solver like \soft{lingeling}) can also impact software variability \cite{xu2008,10.1145/3442391.3442402}.
% Example of this limit: to sharpen (refactor)
The \soft{x264} encoder typifies this problem, as illustrated in \Cref{fig:intro}.
%\fixme{replace the user with devs?}
For example, Kate, an engineer working for a VOD company, wants \soft{x264} to compress input videos to the smallest possible \performance{size}.
She executes \soft{x264} with two configurations C (with options \mbox{-}\mbox{-}\feature{no-mbtree} \mbox{-}\mbox{-}\feature{ref}~\num{1}) and C' (with options \mbox{-}\mbox{-}\feature{no-cabac} \mbox{-}\mbox{-}\feature{ref}~\num{16}) on the input video $I_{1}$ and states that C is more appropriate than C' in this case.
But when trying it on a second input video $I_{2}$, she draws opposite conclusions; for $I_{2}$, C' leads to a smaller output \performance{size} than C. 
Now, Kate wonders 
%Will this performance model predict an accurate value for the \performance{bitrate}? 
what configuration to choose for other inputs, C or C'?
More generally, do configuration options have the same effect on the output \performance{size} despite a different input? 
Do options interact in the same way no matter the inputs? 
These are crucial practical issues: the diversity of existing inputs can alter her knowledge of \soft{x264}'s variability. 
If it does, Kate would have to configure \soft{x264} as many times as there are inputs, making her work really tedious and difficult to automate for a field deployment. 
% measure, the worst situation being to
% basically useless

Numerous research works have proposed to model performance of software configurations, with several use-cases in mind for developers and users of software systems: the maintenance and understanding of configuration options and their interactions \cite{SGKA:ESECFSE15}, the selection of an optimal configuration (tuning) \cite{10.1145/3106237.3106273,FSE2017menzies,flash_find_config}, the performance prediction of arbitrary configurations~\cite{10.1145/3358960.3379137,DBLP:journals/software/KalteneckerGSA20,julianasurvey} or the automated specialization of configurable systems \cite{temple:hal-01467299,IEEEcontextTemple}.
These works measure the performance of several configurations (a sample) under specific settings to then build a performance model. % capable of predicting the performance of any other configuration, \ie a performance model.
Input data further challenges these use-cases, since both the software configuration and the input spaces should be handled. % the combinatorial explosion of 
Inputs also question and can threaten the generalization of configuration knowledge \eg a performance prediction model for a given input may well be meaningless and inaccurate for another input.

On the one hand, some works have been addressing the performance analysis of software systems \cite{blockchain_platform, Coppa2014, image_sensitivity, Goldsmith2007, leitner2016, DBLP:conf/ssbse/SinhaCC20} depending on different input data (also called workloads or benchmarks), but all of them only considered a rather limited set of configurations. 
On the other hand, works and studies on configurable systems usually neglect input data (\eg using a unique video for measuring the configurations of a video encoder). 
We aim to combine both dimensions by performing an in-depth, controlled study of several configurable systems to make it vary in the large, both in terms of configurations and inputs. 
In contrast to research papers mixing multiple factors of the executing environment~\cite{jamshidi2018, DBLP:conf/wosp/ValovPGFC17}, we concentrate on inputs and software configurations only, which allow us to draw reliable conclusions regarding the specific impact of inputs on software variability. 

% ; since inputs affect software performance, it is yet a challenge to train reusable performance prediction models \ie that we could apply on multiple inputs.

%%%%%% CONTRIBS
% Contrib 1: empirical study 
In this paper, we conduct, to our best knowledge, the first in-depth empirical study that measures how inputs individually interacts with software variability. 
To do so, we systematically explore the impact of inputs and configuration options on the performance properties of \nbsystems software systems. 
% Contrib 1: results
This study reveals that inputs fed to software systems can indeed interact with their options in non-monotonous ways, thus significantly impacting their performance properties. 
This observation questions the applicability of performance predictive models trained on only one input: are they still useful for other inputs?
%We provide guidelines to overcome this input sensitivity issue depending on the nature of these interactions. 
We then survey state-of-the-art papers on configurable systems to assess whether they address this kind of input sensitivity issue. 
% on performance prediction
%We found \nbsolutionpapers references detailing a potential solution to the input sensitivity. 
%We found that even when some of them acknowledge the issue, most of them do not mention or address it, which questions their practical relevance. 
%%%%%% CONTRIBS bullet points
Our contributions are as follows: 
\begin{itemize}[noitemsep,leftmargin=0.2cm]
    \item To our best knowledge, the \textbf{first in-depth empirical study that investigates the interactions between input data and configurations} of \nbsystems software systems over \num{1976025} measurements;
    \item We show that \textbf{inputs fed to software systems can interact with their configuration options in non monotonous ways}, thus changing performance of software systems and making it difficult to (automatically) configure them;
    % \item The \textbf{definition of a score of input sensitivity} quantifying the interactions between inputs and configurations;
    \item The \textbf{quantification of input sensitivity} through several indicators and metrics per system and performance property; % about the interplay between inputs and configurations;
    \item An \textbf{analysis of} how \textbf{\nbpapers state-of-the-art research papers} on configurable systems address this problem in practice;
    \item A discussion on the \textbf{impacts of our study} (including key insights and open problems) for different engineering tasks of configurable systems (tuning, prediction, understanding, testing, \etc);
% For each task, we systematically discuss  brought by our results and not addressed in the state of the art.}
    \item \textbf{Open science}: a replication bundle that contains docker images, produced datasets of measurements and code\footnote{Available on Github:\\\scriptsize \bluerl{https://github.com/llesoil/input_sensitivity/tree/master/}}.
\end{itemize}

%\footnote{Available on dockerhub: see \Cref{tab:xp_systems}.}

%%%%%% PAPER PLAN
The remainder of this paper is organized as follows: 
\Cref{sec:problem} explains the problem of input sensitivity and the research questions addressed in this paper. 
\Cref{sec:protocol} presents the experimental protocol. 
\Cref{sec:results} details the results. 
\Cref{sec:significance} shows how researchers address input sensitivity. 
\Cref{sec:discuss} discusses the implications of our work. 
\Cref{sec:threats} details threats to validity. 
% \Cref{sec:related_work} presents related work. 
\Cref{sec:conclusion} summarizes key insights of our paper. 

%%%%% Convention
\smallskip
\textbf{Typographic Convention.} For this paper, we adopt the following typographic convention:
\soft{emphasized} will be relative to a software system, 
\feature{slanted} to its configuration options and 
\performance{underlined} to its performance properties. 

\section{Problem Statement} 
\label{sec:problem}

\subsection{Sensitivity to Inputs of Configurable Systems}
\label{sub:problem:sensitivity}

%%%%%%%%%%%%%%%%% new section 2.1

%%% new context

Configuration options of software systems can have different effects on performance (\eg runtime), but so can the input data. %
For example, a configurable video encoder like \soft{x264} can process many kinds of inputs (videos) in addition to offering options on how to encode. 
Our hypothesis is that there is an interplay between configuration options and input data: some (combinations of) options may have different effects on performance depending on input.

Researchers observed input sensitivity in multiple fields, such as SAT solvers \cite{xu2008, Falkner2015SpySMACAC}, compilation \cite{inputs_compilation,ding2015}, video encoding \cite{Maxiaguine2004}, data compression \cite{8820983}. 
However, existing studies either consider a limited set of configurations (\eg only default configurations), a limited set of performance properties, or a limited set of inputs \cite{netflix-study, blockchain_platform, Coppa2014, image_sensitivity, Goldsmith2007, leitner2016, DBLP:conf/ssbse/SinhaCC20}. 
It limits some key insights about the input sensitivity of configurable systems. 
% Most of the studies support learning models restrictive to specific static settings (\eg inputs, hardware, and version) such that a new prediction model has to be learned from scratch once the environment change \cite{julianasurvey}. 
%The study of Valov \etal \cite{DBLP:conf/wosp/ValovPGFC17} suggests that changing the hardware has reasonable impacts since linear functions are highly accurate when reusing prediction models.
Valov \etal \cite{DBLP:conf/wosp/ValovPGFC17} studied the impact of hardware on software configurations, but fixed the input fed to software systems. 
Jamshidi \etal \cite{jamshidi2017b} explored how environmental conditions (hardware, input, and software versions) impact performances of software configurations. 
Besides considering a limited set of inputs (\eg 3 input videos for \soft{x264}), their study did not aim to isolate the individual effects of input data on software configurations. 
As a result, it is impossible to draw reliable conclusions about the specific variability factors -- among hardware, inputs and versions. 

% contrib
This work details, to the best of our knowledge, the \textbf{first systematic empirical study that analyzes the interactions between input data and configuration options} for different configurable systems. 
Through four research questions introduced in the next section, we characterise the input sensitivity problem and explore how this can alter our understanding of software variability.

%%%%%%%%%%%%%%%%%%%%%%    RQS      %%%%%%%%%%%%%%%%%%%%%%%%%%%%%%%%%%%%%%

\subsection{Research Questions}
\label{sub:problem:rq}

When a developer provides a default configuration for its software system, one should ensure it will perform at best for a large panel of inputs. 
That is, this configuration will be near-optimal whatever the input. 
Hence, a hidden assumption is that two performance distributions over two different inputs are somehow related and close. 
In its simplest form, there could be a linear relationship between these two distributions: they simply increase or decrease with each other.
\textbf{RQ$_{\mathbf{1}}$ - \rqcorr}
To answer this, we compute and compare performance distributions of different inputs. 
For software systems, unstable performance distributions across inputs induce that their optimal configuration change with their inputs. 
In particular, the default configuration should be adapted according to their input data.

But configuration options influence software performance, \eg the energy \performance{consumption} \cite{10.1145/3106195.3106214}. 
An option is called influential for a performance when its values have a strong effect on this performance \cite{jamshidi2018, Dorn2020}. 
For example, developers might wonder whether the option they add to a configurable system has an influence on its performance. 
However, is an option identified as influential for some inputs still influential for other inputs?
If not, it would become both tedious and time-consuming to find influential options on a per-input basis. 
Besides, it is unclear whether activating an option is always worth it in terms of performance; an option could improve the overall performance while reducing it for few inputs. 
If so, users may wonder which options to enable to improve software performance based on their input data. 
\textbf{RQ$_{\mathbf{2}}$ - \rqfeature} 
In this question, we quantify how the effects and importance of software options change with input data. 
If this change is significant, tuning these options to optimize performance should be adapted to the current input.  

$RQ_{1}$ and $RQ_{2}$ study how inputs affect (1) performance distributions and (2) the effects of different configuration options. 
However, the performance distributions could change in a negligible way, without affecting the software user's experience. 
Before concluding on the real impact of the input sensitivity, it is necessary to quantify how much this performance changes from one input to another. 
\textbf{RQ$_{\mathbf{3}}$ - \rqimpact} 
In particular, we estimate the loss in performance when configuring a software while ignoring the input sensitivity to inputs. 
To put it more positively, this loss is also the potential gain, in terms of performance, to tune a software system for its input data. 

$RQ_{1}$ to $RQ_{3}$ present the problem of input sensitivity. 
The fourth and last question explores the limits of this problem and give insights on how to address it concretely. 
Though all inputs are different, the number of possible interactions between the software systems and the processed inputs is limited. 
Therefore, there might exist inputs interacting in the same way with the software, and thus having similar performance profiles. 
\textbf{RQ$_{\mathbf{4}}$ - \rqgroup} 
For this question, we form, analyze and characterize different groups of inputs having similar performance distributions and show the benefits of these groups to address the input sensitivity issue. 

\section{Experimental protocol}
\label{sec:protocol}

We designed the following experimental protocol to answer these research questions.

\begin{table*}[t]
% \vspace*{-0.3cm}
\centering
\caption{Subject Systems. See \Cref{fig:measurement_process} for notations.}
\label{tab:xp_systems}
% \vspace*{-0.3cm}

\renewcommand{\arraystretch}{0.8}
\setlength{\tabcolsep}{2.2pt} % 1.8
% \hspace*{-2.5cm}
\begin{tabular}{|c|c|c|c|c|c|c|c|c|c|c|}
\hline
\scriptsize \cellcolor[HTML]{e8e8e8}{\textit{\textbf{System}}}
 &\scriptsize \cellcolor[HTML]{e8e8e8}{\textit{\textbf{Domain}}}
 &\scriptsize \cellcolor[HTML]{e8e8e8}{\textit{\textbf{Commit}}}  
 &\scriptsize \cellcolor[HTML]{e8e8e8}{\textit{\textbf{\#LoCs}}}
 &\scriptsize \cellcolor[HTML]{acc9fa}{\textit{\textbf{Configs \#C}}}
 &\scriptsize \cellcolor[HTML]{ea9e99}{\textit{\textbf{Inputs I}}}
 &\scriptsize \cellcolor[HTML]{ea9e99}{\textit{\textbf{\#I}}}
 &\scriptsize \cellcolor[HTML]{b3e6b3}{\textit{\textbf{\#M}}}
 &\scriptsize \cellcolor[HTML]{b3e6b3}{\textit{\textbf{Performance(s) P}}}
 &\scriptsize \cellcolor[HTML]{ffffff}{\textit{\textbf{Docker}}}
 &\scriptsize \cellcolor[HTML]{ffffff}{\textit{\textbf{Dataset}}} \\ 
 \hline  
 
 % gcc
\scriptsize \textcolor{blue}{\href{https://gcc.gnu.org/}{gcc}} 
&\scriptsize Compilation 
&\scriptsize \textcolor{blue}{\href{https://github.com/gcc-mirror/gcc/commit/ccb4e0774b3e5859ea1d7f1864b02fa5826c4a79}{ccb4e07}} 
&\scriptsize \num{9606697}
&\scriptsize \num{80} 
&\scriptsize \emph{.c} programs 
&\scriptsize \num{30} 
&\scriptsize \num{2400} 
&\scriptsize size, ctime, exec 
&\scriptsize \textcolor{blue}{\href{https://hub.docker.com/r/anonymicse2021/gcc_inputs}{Link}} 
&\scriptsize \textcolor{blue}{\href{https://zenodo.org/record/5136613}{Link}} \tabularnewline \hline 

% ImageMagick
\scriptsize \textcolor{blue}{\href{https://imagemagick.org/index.php}{ImageMagick}} 
&\scriptsize Image processing 
&\scriptsize \textcolor{blue}{\href{https://github.com/ImageMagick/ImageMagick/commit/5ee49d66e6534ab7d145dce89e502a6d0b9f18fa}{5ee49d6}} 
&\scriptsize \num{648984}
&\scriptsize \num{100} 
&\scriptsize images 
&\scriptsize \num{1000} 
& \scriptsize \num{100000} 
& \scriptsize size, time 
&\scriptsize \textcolor{blue}{\href{https://hub.docker.com/r/anonymicse2021/imagemagick_inputs}{Link}}
&\scriptsize \textcolor{blue}{\href{https://zenodo.org/record/5145853}{Link}} \tabularnewline \hline

% lingeling (solveur sat)
\scriptsize \textcolor{blue}{\href{http://fmv.jku.at/lingeling/}{lingeling}} 
&\scriptsize SAT solver 
&\scriptsize \textcolor{blue}{\href{https://github.com/arminbiere/lingeling/commit/7d5db72420b95ab356c98ca7f7a4681ed2c59c70}{7d5db72}} 
&\scriptsize \num{33402}
&\scriptsize \num{100} 
&\scriptsize SAT formulae 
&\scriptsize \num{351} 
& \scriptsize \num{35100} 
&\scriptsize \#confl.,\#reduc. 
&\scriptsize \textcolor{blue}{\href{https://hub.docker.com/r/anonymicse2021/lingeling_inputs}{Link}} 
&\scriptsize \textcolor{blue}{\href{https://zenodo.org/record/5101310}{Link}} \tabularnewline \hline  

% nodejs
\scriptsize \textcolor{blue}{\href{https://nodejs.org/en/}{nodeJS}} 
&\scriptsize JS runtime env. 
&\scriptsize \textcolor{blue}{\href{https://github.com/nodejs/node/commit/78343bbdc572590886c9da53a73b6061e62a5f3e}{78343bb}} 
&\scriptsize \num{6205618}
&\scriptsize \num{50} 
&\scriptsize \emph{.js} scripts 
&\scriptsize \num{1939} 
&\scriptsize \num{96950} 
&\scriptsize \#operations/s 
&\scriptsize \textcolor{blue}{\href{https://hub.docker.com/r/anonymicse2021/nodejs_inputs}{Link}}
&\scriptsize \textcolor{blue}{\href{https://zenodo.org/record/5067851}{Link}} \tabularnewline \hline  

% poppler
\scriptsize \textcolor{blue}{\href{https://poppler.freedesktop.org}{poppler}} 
&\scriptsize PDF rendering  
&\scriptsize \textcolor{blue}{\href{https://github.com/freedesktop/poppler/commit/42dde686bf5a674401850b2d3fdd2bc7467e9a66}{42dde68}} 
&\scriptsize \num{197019}
&\scriptsize \num{16} 
&\scriptsize \emph{.pdf} files 
&\scriptsize \num{1480} 
&\scriptsize \num{23680} 
&\scriptsize size, time 
&\scriptsize \textcolor{blue}{\href{https://hub.docker.com/r/anonymicse2021/poppler_inputs}{Link}}
&\scriptsize \textcolor{blue}{\href{https://zenodo.org/record/5033478}{Link}} \tabularnewline \hline 

% SQLite
\scriptsize \textcolor{blue}{\href{https://sqlite.org/index.html}{SQLite}} 
&\scriptsize DBMS 
&\scriptsize \textcolor{blue}{\href{https://github.com/sqlite/sqlite/commit/53fa02507b2025db7b74a155c8df4a8a2e4db4d8}{53fa025}} 
&\scriptsize \num{320049}
&\scriptsize \num{50} 
&\scriptsize databases 
&\scriptsize \num{150} 
&\scriptsize \num{7500}  
&\scriptsize \num{15} query times q\num{1}-q\num{15} 
& \scriptsize \textcolor{blue}{\href{https://hub.docker.com/r/anonymicse2021/sqlite_inputs}{Link}}
&\scriptsize \textcolor{blue}{\href{https://zenodo.org/record/5139331}{Link}} \tabularnewline \hline  

% x264
\scriptsize \textcolor{blue}{\href{https://www.videolan.org/developers/x264.html}{x264}} 
&\scriptsize Video encoding 
& \scriptsize \textcolor{blue}{\href{https://github.com/mirror/x264/commit/e9a5903edf8ca59ef20e6f4894c196f135af735e}{e9a5903}} 
&\scriptsize \num{110642}
&\scriptsize \num{201} 
&\scriptsize videos 
&\scriptsize \num{1397} 
&\scriptsize \num{280797} 
&\scriptsize cpu, fps, kbs, size, time 
&\scriptsize \textcolor{blue}{\href{https://hub.docker.com/r/anonymicse2021/x264_inputs}{Link}} 
&\scriptsize \textcolor{blue}{\href{https://zenodo.org/record/3928253}{Link}} \tabularnewline \hline 

% xz
\scriptsize \textcolor{blue}{\href{https://tukaani.org/xz/}{xz}} 
&\scriptsize Data compression 
&\scriptsize \textcolor{blue}{\href{https://github.com/xz-mirror/xz/commit/e7da44d5151e21f153925781ad29334ae0786101}{e7da44d}} 
&\scriptsize \num{37489}
&\scriptsize \num{30} 
&\scriptsize system files 
&\scriptsize \num{48} 
&\scriptsize \num{1440} 
&\scriptsize size, time 
&\scriptsize \textcolor{blue}{\href{https://hub.docker.com/r/anonymicse2021/xz_inputs}{Link}} 
&\scriptsize \textcolor{blue}{\href{https://zenodo.org/record/5033489}{Link}} \tabularnewline \hline  
\end{tabular}
%\vspace*{-0.2cm}
\end{table*}

\begin{figure}[ht]
\centering
% \vspace*{-0.2cm}
\includegraphics[width=0.75\columnwidth]{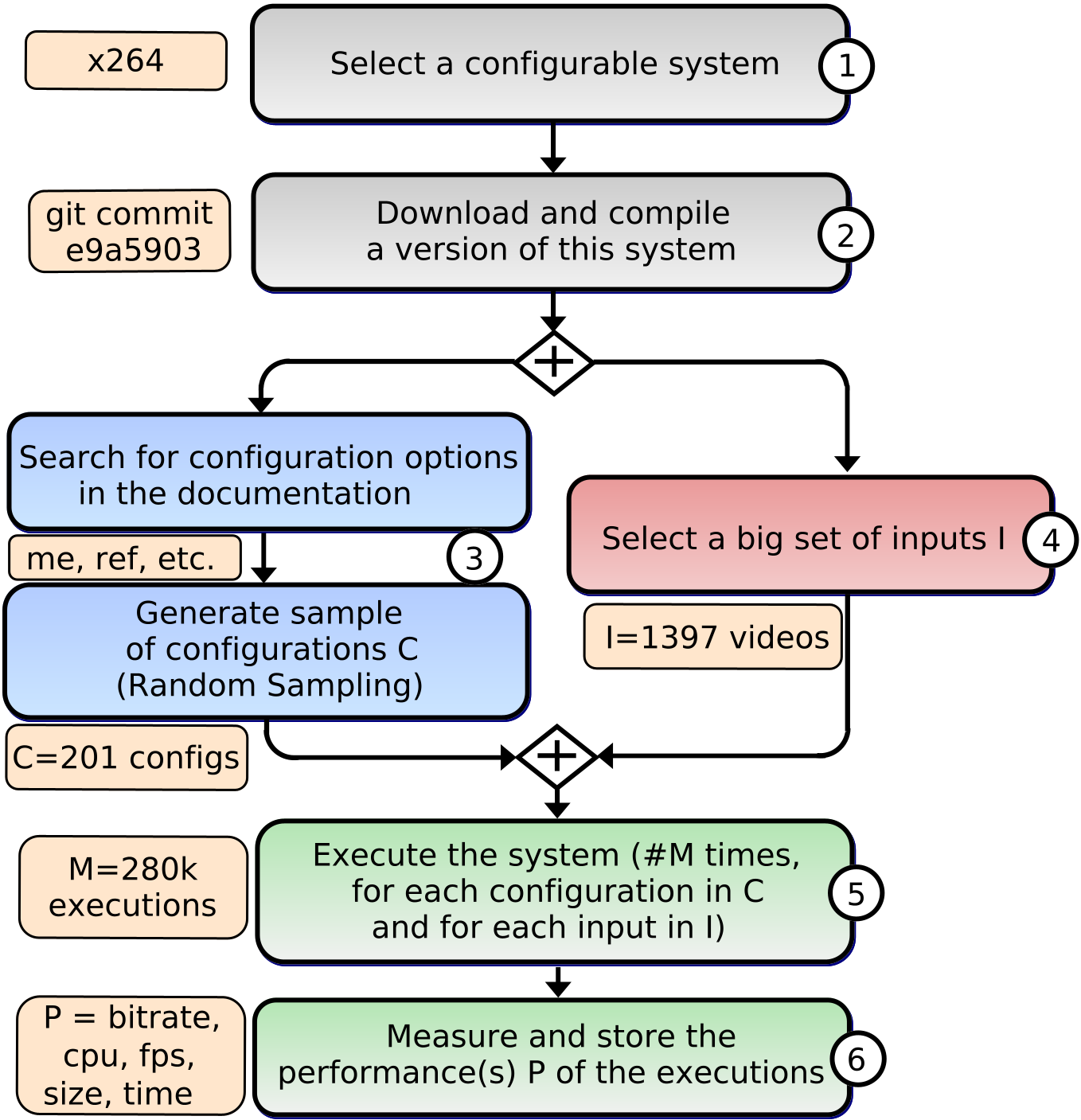}
% \vspace*{-0.3cm}
\caption{Measuring performance - Protocol}
\label{fig:measurement_process}
% \vspace*{-0.4cm}
\end{figure}
%input{figure/protocol-table_systems}

\subsection{Data Collection}
\label{sub:protocol:measuring}

We first collect measurements of systems processing inputs. 

%\smallskip
\textbf{Protocol.} \Cref{fig:measurement_process} depicts the step-by-step protocol we respect to measure performance of software systems. Each line of \Cref{tab:xp_systems} should be read following \Cref{fig:measurement_process}: System with Steps \num{1} and \num{2}; Configurations \#C with Step \num{3}; the nature of inputs I and their number \#I with Step \num{4}; Performance P with Steps \num{5} and \num{6}; Docker links a container for executing all the steps and Data the datasets containing the performance measurements. 
\Cref{fig:measurement_process} shows in beige an example with the \soft{x264} encoder. 
Hereafter, we provide details for each step of the protocol.

%\smallskip
\textbf{Steps 1 \& 2 - Software Systems.} 
We consider \nbsystems software systems. We choose them because they are open-source, well-known in various fields and already studied in the literature: \soft{gcc} \cite{inputs_compilation,9401979}, the compiler for gnu operating system; \soft{ImageMagick} \cite{stillimagemagick,smitha2016document}, a software system processing pictures and images; \soft{lingeling} \cite{heule2018proceedings,Falkner2015SpySMACAC}, a SAT solver; \soft{nodeJS} \cite{incerto2017software,10.1145/3178372.3179527}, a widely-used JavaScript execution environment; \soft{poppler} \cite{kilgard2020anecdotal,8674171}, a library designed to process \emph{.pdf} files; \soft{SQLite}~\cite{IEEEcontextTemple,jamshidi2017b}, a database manager system; \soft{x264} \cite{jamshidi2017b,10.1145/3358960.3379137}, a video encoder based on H264 specifications; \soft{xz} \cite{DBLP:conf/wosp/ValovPGFC17,9285664}, a file system manager. 
We also choose these systems because they handle different types of input data, allowing us to draw conclusions as general as possible. 
For each software system, we use a unique private server with the same configuration running over the same operating system\footnote{The configurations of the running environments are available at: \scriptsize{\bluerl{https://github.com/llesoil/input_sensitivity/tree/master/replication/Environments.md}}}. 
We download and compile a unique version of the system. All performance are measured with this version of the software. 

%\smallskip
\textbf{Step 3 - Configuration options C.} 
To select the configuration options, we read the documentation of each system and manually extract the options affecting the performance of the system. For instance, according to the documentation of \soft{x264}, the option \mbox{-}\mbox{-}\feature{mbtree} "\textit{can lead to large savings for very flat content}" and "\textit{animated content should use stronger} \mbox{-}\mbox{-}\feature{deblock} \textit{settings}".\footnote{See the documentation of \soft{x264} at \scriptsize{\bluerl{https://silentaperture.gitlab.io/mdbook-guide/encoding/x264.html}}} 
Out of these configuration options, we then sample \#C configurations by using random sampling~\cite{8730148}. 
In the previous example, after the selection of \mbox{-}\mbox{-}\feature{mbtree} and \mbox{-}\mbox{-}\feature{deblock}, the sampling step would generate multiple configurations with combinations of options' values: $C_{1}$, with \mbox{-}\mbox{-}\feature{mbtree} activated and \mbox{-}\mbox{-}\feature{deblock} set to "0:0"; $C_{2}$, with \mbox{-}\mbox{-}\feature{mbtree} deactivated and \mbox{-}\mbox{-}\feature{deblock} set to "-2:-2";
$C_{3}$, with \mbox{-}\mbox{-}\feature{mbtree} deactivated and \mbox{-}\mbox{-}\feature{deblock} set to "0:0". 
To ensure that each value of a software option is well represented in the final set of configurations, we statistically test the uniformity of its values. 
To do so, we apply a Kolmogorov-Smirnov test~\cite{massey1951kolmogorov} to each option of our eight software systems\footnote{Options and tests results are available at: \scriptsize{\bluerl{https://github.com/llesoil/input_sensitivity/tree/master/results/others/configs/sampling.md}}}.  
In the previous example, for a boolean option like \mbox{-}\mbox{-}\feature{mbtree} that can be either activated or deactivated, a valid Kolmogorov-Smirnov test guarantees that \mbox{-}\mbox{-}\feature{mbtree} is activated in roughly \pc{50} of the configurations. 
To mitigate the threat of only using random sampling, we also considered various informed configurations picked in the documentation. For instance, for \soft{x264}, we considered the ten presets configurations recommended by the documentation\footnote{See \url{http://www.chaneru.com/Roku/HLS/X264_Settings.htm\#preset}}

\textbf{Step 4 - Inputs I.}
For each system, we select a different set of input data: for \soft{gcc}, PolyBench v\num{3.1} \cite{pouchet2012polybench}; for \soft{ImageMagick}, a sample of ImageNet \cite{5206848} images (from \kb{1.1} to \mb{7.3}); for \soft{lingeling}, the 2018 SAT competition's benchmark \cite{heule2018proceedings}; for \soft{nodeJS}, its test suite;  for \soft{poppler}, the Trent Nelson's PDF Collection \cite{tpn_dataset}; for \soft{SQLite}, a set of generated TPC-H \cite{10.1145/369275.369291} databases (from \mb{10} to \gb{6}); for \soft{x264}, the YouTube \textbf{U}ser \textbf{G}eneral \textbf{C}ontent dataset \cite{Wang2019} of videos (from \mb{2.7} to \gb{39.7}); for \soft{xz}, a combination of the Silesia \cite{silesia_corpus} and the Canterbury\cite{canterbury_corpus} corpus.
We choose them because these are large and freely available datasets of inputs, well-known in their field and already used by researchers and practitioners. 

%\smallskip
\textbf{Steps 5 \& 6 - Performance properties P.} For each system, we systematically execute all the configurations of C on all the inputs of I. For the \#M resulting executions, we measure as many performance properties as possible: for \soft{gcc}, \performance{ctime} and \performance{exec} the times needed to compile and execute a program and the \performance{size} of the binary; for \soft{ImageMagick}, the \performance{time} to apply a Gaussian blur \cite{hummel1987deblurring} to an image and the \performance{size} of the resulting image; for \soft{lingeling}, the number of \performance{conflicts}  and \performance{reductions} found in \num{10} seconds of execution; for \soft{nodeJS},  the number of operations per second (\performance{ops}) executed by the script; for \soft{poppler}, the time needed to extract the images of the pdf, and the \performance{size} of the images; for \soft{SQLite}, the time needed to answer \num{15} different queries \performance{q\num{1}}$\to$ \performance{q\num{15}}; for \soft{x264}, the \performance{bitrate} (the average amount of data encoded per second), the \performance{cpu} usage (percentage), the average number of frames encoded per second (\performance{fps}), the \performance{size} of the compressed video and the elapsed \performance{time}; for \soft{xz}, the \performance{size} of the compressed file, and the \performance{time} needed to compress it. It results in a set a tabular data, one for each input and each software system, consisting of a list of configurations with their performance property values. 

%\smallskip
\textbf{Replication.}
To allow researchers to easily replicate the measurement process, we provide a docker container for each system (see the links in the \textit{Docker} column of \Cref{tab:xp_systems}). 
We also publish the resulting datasets online (see the links in the \textit{Data} column) and in the companion repository with replication details\footnote{Guidelines for replication are available at: \scriptsize{\bluerl{https://github.com/llesoil/input_sensitivity/tree/master/replication/README.md}}}.

%For the next research questions, our results are computed with Python v\num{3}.\num{7}.\num{6} and specific versions of data science libraries.\footnote{The description of the python environment is available at: \scriptsize{\bluerl{https://anonymous.4open.science/r/df319578-8767-47b0-919d-a8e57eb67d25/replication/requirements.txt}}}

%%%%%%%%%%%%%%%%%%%%%%%%%%%%%%%%%%   RQ1    %%%%%%%%%%%%%%%%%%%%%%%%%%%%%%%
% \vspace*{-0.1cm}

\subsection{Performance Correlations ($RQ_{1}$)}
\label{sub:protocol:corr}

Based on the analysis of the data collected in \Cref{sub:protocol:measuring}, we can now answer the first research question: 
\textbf{RQ$_{\mathbf{1}}$ - \rqcorr}\xspace
To check this hypothesis, we compute, analyze and compare the Spearman's rank-order correlation \cite{spearman} of each couple of inputs for each system. 
It is appropriate in our case since all performance properties are quantitative variables measured on the same set of configurations. 

\textbf{Spearman correlations.} 
The correlations are considered as a measure of similarity between the configurations' performance over two inputs. 
We compute the related $p$-values: a correlation whose $p$-value is higher than the chosen threshold \num{0.05} is considered as null.
We use the Evans rule \cite{evans1996_correlation} to interpret these correlations. In absolute value, we refer to correlations by the following labels; very low: 0-0.19, low: 0.2-0.39, moderate: 0.4-0.59, strong: 0.6-0.79, very strong: 0.8-1.00. 
A negative score tends to reverse the ranking of configurations. 
Very low or negative scores have practical implications: a good configuration for an input can very well exhibit bad performance for another input.

%%%%%%%%%%%%%%%%%%%%%%%%%%%%%%%%%%%%%%%%%%%%%%%%%%%%%%  RQ2    %%%%%%%%%%%%%%%%%%%%%%%%%%%%%%%
% \vspace*{-0.1cm}

\subsection{Effects of Options ($RQ_{2}$)}
\label{sub:protocol:feature}

To understand how a performance model can change based on a given input, we next study how input data interact with configuration options. 
\textbf{RQ$_{\mathbf{2}}$ - \rqfeature}\xspace
To assess the relative significance and effect of options, we use two well-known statistical methods \cite{linear_reg, breiman2001random}, also widely used in the context of interpretable machine learning and configurable systems~\cite{julianasurvey,molnar2019, jamshidi2017b}.
For instance, Jamshidi \etal \cite{jamshidi2017b} used similar indicators to measure the sensitivity of configurations regarding computing environmental conditions (hardware, input, and software versions).

% the importance of features vary with inputs, i.e. how they are important to predict the performance, regardless of the effect 
\textbf{Random forest importances.} The tree structure provides insights about the most essential options for prediction, because such a tree first splits \wrt options that provide the highest information gain. 
We use random forests \cite{breiman2001random}, a vote between multiple decision trees:
we can derive, from the forests trained on the inputs, estimates of the options importance. 
The computation of option importance is realized through the observation of the eﬀect on random forest accuracy when randomly shuffling each predictor variable \cite{molnar2019}.
For a random forest, we consider that an option is influential if the median (on all inputs) of its option importance is greater than $\frac{1}{n_{opt}}$, where $n_{opt}$ is the number of options considered in the dataset. 
This threshold represents the theoretic importance of options for a software having equally important options.% -inspired by the Kaiser rule \cite{kaiser_rule}. 

% a feature can increase or diminish the performance depending on the input, here we measure the effect
\textbf{Linear regression coefficients.} The coefficients of an ordinary least square regression \cite{linear_reg} weight the effect of configuration options. 
These coefficients can be positive (\resp negative) if a bigger (\resp lower) option value results in a bigger performance. 
Ideally, the sign of the coefficients of a given option should remain the same for all inputs: it would suggest that the effect of an option onto performance is stable.
We also provide details about coefficients related to feature interactions \cite{DBLP:conf/splc/ValovGC15,guo2015} in the companion repository.

%%%%%%%%%%%%%%%%%%%%%%%%%%%%%%%%%%%%%%%%%%%%%%%%%%%%%%  RQ3 %%%%%%%%%%%%%%%%%%%%%%%%%%%%%%%

\subsection{Impact of Inputs on Performance ($RQ_{3}$)}
\label{sub:protocol:impact}

To complete this experimental protocol, we ask whether adapting the software to its input data is worth the cost of finding the right set of parameters \ie the concrete impact of input sensitivity. 
\textbf{RQ$_{\mathbf{3}}$ - \rqimpact}\xspace
To estimate how much we can lose, we first define two scenarios of reuse $S_{1}$ and $S_{2}$: 

\begin{enumerate}[label=$S_{\arabic*}$ -,topsep=0pt,itemsep=-1ex,partopsep=1ex,parsep=1ex, leftmargin=0.6cm]
    \item \textit{Baseline.} In this scenario, we value input sensitivity and just train a simple performance model on a target input. We choose the best configuration according to the model, configure the related software with it and execute it on the target input.
    \item \textit{Ignoring input sensitivity.} In this scenario, let us pretend that we ignore the input sensitivity issue. We train a model related to a given input \ie the source input, and then predict the best configuration for this source input. If we ignore the issue of input sensitivity, we should be able to easily reuse this model for any other input, including the target input of $S_{1}$. Finally, we execute the software with the predicted configuration on the target input.
\end{enumerate}

In this part, we systematically compare $S_{1}$ and $S_{2}$ in terms of performance for all inputs, all performance properties and all software systems. 
For $S_{1}$, we repeat the scenario ten times with different sources, uniformly chosen among other inputs and compute the average performance. 
For both scenarios, due to the imprecision of the learning procedure, the models can recommend sub-optimal configurations. 
Since this imprecision can alter the results, we consider an ideal case for both scenarios and assume that the performance models always recommend the best possible configuration. 

%To avoid adding this imprecision to the effect of input sensitivity, and in order to be fair, we consider that the models are oracles \ie that they predict the best configuration each time.

\textbf{Performance ratio.}
To compare $S_{1}$ and $S_{2}$, we use a performance ratio \ie the performance obtained in $S_{1}$ over the performance obtained in $S_{2}$. 
If the ratio is equal to \num{1}, there is no difference between $S_{1}$ and $S_{2}$ and the input sensitivity does not exist.
A ratio of \num{1.4} would suggest that the performance of $S_{1}$ is worth \num{1.4} times the performance of $S_{2}$; therefore, it is possible to gain up to $(1.4-1)*100=40\%$ performance by choosing $S_{1}$ instead of $S_{2}$. 
We also report on the standard deviation of the performance ratio distribution. 
A standard deviation of 0 implies that we gain or lose the same proportion of performance when picking $S_{1}$ over $S_{2}$. 
% As a comparison, we compute the performance ratio between extreme configurations \ie the best over the worst.
%\fixme{s'il reste de la place, un petit test pour prouver que $S_{1}>S_{2}$?}

%%%%%%%%%%%%%%%%%%%%%%%%%%%%%%%%%%%%%%%%%%%%%%%%%%%%%%  RQ4  %%%%%%%%%%%%%%%%%%%%%%%%%%%%%%%

\subsection{Groups of Inputs ($RQ_{4}$)}
\label{sub:protocol:groups}

Lastly, we explore how the issue of input sensitivity can be concretely addressed.
For mitigating input sensitivity, an idea is to group together inputs based on their performance distributions. 
\textbf{RQ$_{\mathbf{4}}$ - \rqgroup}\xspace
The inputs belonging to the same group are supposed to share common properties and be processed in a similar manner by the software~\cite{ding2015}. 
We perform hierarchical clustering~\cite{nielsen2016hierarchical} to gather inputs having similar performance profiles.

\textbf{Hierarchical clustering.} 
This technique considers the correlations between performance distributions as a measure of similarity between inputs. Based on these values, it forms groups of inputs minimizing the intra-class variance ( discrepancy of performance among a group) and maximizing the inter-class variance (discrepancy of performance between different groups of inputs).
As linkage criteria, we choose the Ward method~\cite{nielsen2016hierarchical} since in our case, (1) single link (minimum of distance) leads to numerous tiny groups (2) centroid or average tend to split homogeneous groups of inputs and (3) complete link aggregates unbalanced groups.  
As a metric, we kept the Euclidian distance - used as default. 
We manually select the final number of groups. 

For each group, we then report on few key indicators summarizing the specifics of inputs' performance: the Spearman correlations between performance distributions of inputs ($RQ_{1}$), the importance and effects of options ($RQ_{2}$) as well as few properties characterizing the inputs \eg the spatial complexity of an input video or the number of lines of a \emph{.c} program. 
We compare their average value in the different groups. 

\section{Results}
\label{sec:results}

%\noindent We now present the results obtained by following the methodology defined in \Cref{sec:protocol}.
%: \Cref{sub:results:corr} \wrt \Cref{sub:protocol:corr}, \Cref{sub:results:feature} \wrt \Cref{sub:protocol:feature}, \Cref{sub:results:impact} \wrt \Cref{sub:protocol:impact} and \Cref{sub:results:papers} \wrt \Cref{sub:protocol:papers}.

%%%%%%%%%%%%%%%%%%%%%%%%%%%%%%%%%%%%%% RQ1 %%%%%%%%%%%%%%%%%%%%%%%%%%%%%%%%%%%%%%%%%%%%%%%%%%%

\subsection{Performance Correlations ($RQ_{1}$)} 
\label{sub:results:corr}

We first explain the results of $RQ_{1}$ and their consequences on the \soft{poppler} use case \ie an extreme case of input sensitivity, and then generalize to our other software systems. 

\begin{figure}[t]
% \vspace*{-0.2cm}
\includegraphics[width=\columnwidth]{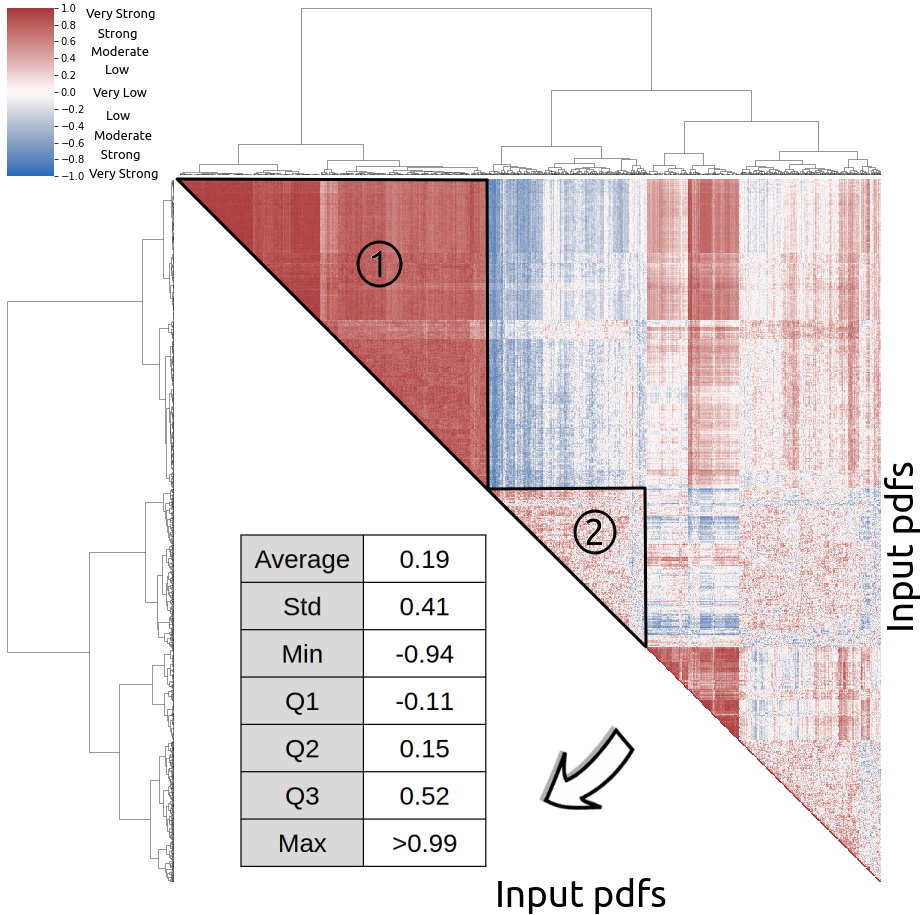}
\footnotesize
Each square$_{(i,j)}$ represents the Spearman correlation between the \performance{time} needed to extract the images of pdfs $i$ and $j$. The color of this square respects the top-left scale: high positive correlations are red; low in white; negative in blue. Because we cannot describe each correlation individually, we added a table describing their distribution. %(diagonal excluded).
% \vspace*{-0.3cm}
\caption{Spearman rank based correlations - \soft{poppler}, \performance{time}.}
\label{fig:spearmansize}
% \vspace*{-0.3cm}
\end{figure}

%% use case : poppler 
\textbf{Extract images of input pdfs with \soft{poppler}.} 
The content of pdf files fed to \soft{poppler} may vary; the input pdf can contain a \num{2}-page extended abstract with plain text, a \num{10}-page conference article with few figures or a \num{300}-page book full of pictures. 
Depending on this content, extracting the images embedded in those files can be quicker or slower for the same configuration. 
Moreover, different configurations could be adapted for the conference paper but not for the book (or conversely), leading to different rankings of extraction \performance{time} and thus different rank-based correlation values. 

%fig + résultats simples
\Cref{fig:spearmansize} depicts the Spearman rank-order correlations of extraction \performance{time} between pairs of input pdfs fed to \soft{poppler}. 
Results suggest a positive correlation (see dark red cells), though there are pairs of inputs with lower (see white cells) and even negative (see dark blue cells) correlations. 
More than a quarter of the correlations between input pdfs are positive and at least moderate - third quartile Q3 greater than \num{0.52}.

\begin{figure*}[!ht]
    \centering
    \begin{subfigure}{\columnwidth}
        \centering
        \includegraphics[width=1.02\linewidth]{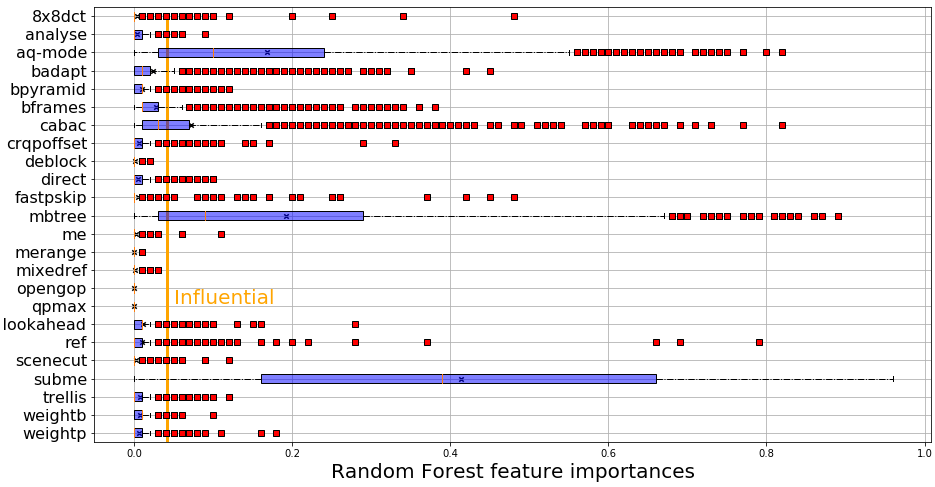}
        % \vspace*{-0.5cm}
        \caption{Importance}
        \label{fig:featurerfboxplot-kbs}
    \end{subfigure}
    % \hspace{0.1cm}
    \begin{subfigure}{\columnwidth}
        \centering
        \includegraphics[width=1.02\linewidth]{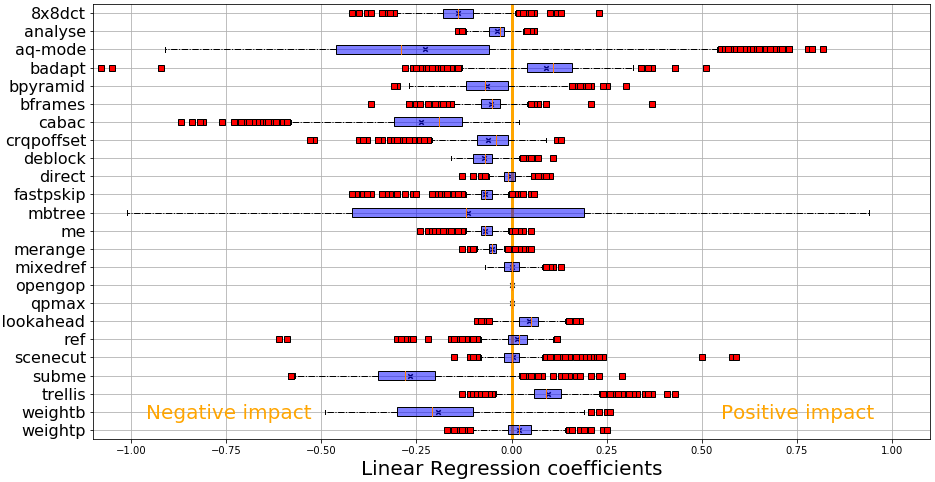}
        % \vspace*{-0.5cm}
        \caption{Effect}
        \label{fig:featurelinboxplot-kbs}
    \end{subfigure}
    % \vspace*{-0.3cm}
    \caption{Importance and effect (x-axis) of configuration options (y-axis) - \soft{x264}, \performance{bitrate}}
    \label{fig:featurepoly-coeff-kbs} 
    % \vspace*{-0.3cm}
\end{figure*}

%% Généralisation
\textbf{Meta-analysis.}
Over the \nbsystems systems\footnote{Detailed $RQ_{1}$ results for other systems are available at: \scriptsize{\bluerl{https://github.com/llesoil/input_sensitivity/tree/master/results/RQS/RQ1/RQ1.md}}}, we observe different cases. There exist software systems not sensitive at all to inputs. In our experiment, \soft{gcc}, \soft{imagemagick} and \soft{xz} present almost exclusively high and positive correlations between inputs \eg Q1 = \num{0.82} for the compressed \performance{size} and \soft{xz}. For these, un- or negatively-correlated inputs are an exception more than a rule. In contrast, there are software systems, namely \soft{lingeling}, \soft{nodeJS}, \soft{SQLite} and \soft{poppler}, for which performance distributions completely change and depend on input data \eg Q2 = \num{0.09} for \soft{nodeJS} and \performance{ops}, Q3 = \num{0.12} for \soft{lingeling} and \performance{conflicts}. For these, we draw similar conclusions as in the \soft{poppler} case. In between, \soft{x264} is only input-sensitive \wrt a performance property; it is for \performance{bitrate} and \performance{size} but not for \performance{cpu}, \performance{fps} and \performance{time} \eg \num{0.29} as deviation for \performance{size} against \num{0.08} for \performance{time}.

% \vspace*{-0.05cm}

%% conclusion RQ1
\begin{tcolorbox}[boxsep=-2pt]
\textbf{RQ$_{\mathbf{1}}$ - \rqcorr} 
%Our systematic empirical study 
%Performance distributions change depending on inputs. 
%It has practi
% We show evidences about the existence of input sensitivity: 
% (1) input sensitivity does affect a majority of our systems (5/8), but not all of them; 
% (2) input sensitivity may affect not the whole systems but few performance properties. 
% So, without having scrutinized the input sensitivity of a system, one cannot develop techniques sensitive to this phenomenon.
We show that :
(1) depending on the inputs, the rank-based correlations of performance distribution can be high, close to zero, or even negative; 
(2) since configuration rankings can change with input data, the best configuration for an input will not be the best configuration for another input. 
The consequence is that one cannot blindly reuse a configuration prediction model across inputs and that developers should not provide to end-users a unique default configuration whatever the input is. 
\end{tcolorbox}

%%%%%%%%%%%%%%%%%%%%%%%%%%%%%%%%%%%%%% RQ2 %%%%%%%%%%%%%%%%%%%%%%%%%%%%%%%%%%%%%%%%%%%%%%%%%%%

\subsection{Effects of Options ($RQ_{2}$)} 
\label{sub:results:feature}

We first explain the results of $RQ_{2}$ and their concrete consequences on the \performance{bitrate} of \soft{x264} - an input-sensitive case, to then generalize to other software systems.

%% use case : x264
\textbf{Encoding input videos with \soft{x264}.}
\Cref{fig:featurerfboxplot-kbs,fig:featurelinboxplot-kbs} report on respectively the boxplots of configuration options' feature importances and effects when predicting \soft{x264}'s \performance{bitrate} for all input videos\footnote{Detailed $RQ_{2}$ results for other systems are available at: \scriptsize{\bluerl{https://github.com/llesoil/input_sensitivity/tree/master/results/RQS/RQ2/RQ2.md}}}. 
On the top graph, we displayed the boxplots of the distribution of importances for each option (y-axis). On the bottom graph, we displayed the boxplots of the distribution of regression coefficients for each option (y-axis). Each red square is representing a model trained on one input, and all of them constitute the resulting distribution.
Each algorithm is using \pc{100} of the configurations in the training set. We also compute the Mean Absolute Percentage Error (MAPE) for all the systems, inputs and non functional properties when predicting the performance with random forests and linear regression. For random forest, we can ensure that our models are giving a good prediction, the median value of the MAPE across all inputs being systematically under \pc{5} for each couple of software systems and performance properties. For Linear Regression, results tend to show higher values of MAPE, suggesting that the configurations spaces are too hard to learn from for such  simple models.
Other variants of feature importances and linear regression (permutation importance\footnote{See results at \url{https://github.com/llesoil/input_sensitivity/blob/master/results/RQS/RQ2/RQ2_permutation.ipynb}}, drop-column importance\footnote{See results at \url{https://github.com/llesoil/input_sensitivity/blob/master/results/RQS/RQ2/RQ2_drop.ipynb}} and Shapley values\footnote{See results at \url{https://github.com/llesoil/input_sensitivity/blob/master/results/RQS/RQ2/RQ2_shapley.ipynb}}) have been computed to ensure the robustness of results in the companion repository. They reached similar results, which confirms our conclusions with the chosen indicators.

% results importances x264
Three options are strongly influential for a majority of videos on \Cref{fig:featurerfboxplot-kbs}: \feature{subme}, \feature{mbtree} and \feature{aq-mode}, but their importance can differ depending on input videos: for instance, the importance of \feature{subme} is \num{0.83} for \video{1365} and only \num{0.01} for \video{40}. 
Because influential options vary with inputs, performance models and approaches based on \textit{feature selection}~\cite{molnar2019} such as performance-influence model~\cite{SGKA:ESECFSE15,weber2021whitebox} may not generalize well to all input videos. 

% results effects x264
Most of the options have positive and negative coefficients on \Cref{fig:featurelinboxplot-kbs}; thus, the specific effects of options heavily depend on input videos. 
It is also true for influential options: \feature{mbtree} can have positive and negative (influential) effects on the \performance{bitrate} \ie activating \feature{mbtree} may be worth only for few input videos.
The consequence is that tuning the options of a software system should be adapted to the current input, and not done once for all the inputs. 

%% Généralisation 
\textbf{Meta-analysis.} For \soft{gcc}, \soft{imagemagick} and \soft{xz}, the importances are quite stable. As an extreme case of stability, the importances of the compressed \performance{size} for \soft{xz} are exactly the same, except for two inputs. For these systems, the coefficients of linear regression mostly keep the same sign across inputs \ie the effects of options do not change with inputs. 
For input-sensitive software systems, we always observe high variations of options' effects (\soft{lingeling}, \soft{poppler} or \soft{SQLite}), sometimes coupled to high variations of options' importances (\soft{nodeJS}). For instance, the option \feature{format} for \soft{poppler} can have an importance of \num{0} or \num{1} depending on the input. For all software systems, there exists at least one performance property whose effects are not stable for all inputs \eg one input with negative coefficient and another with a positive coefficient. 
For \soft{x264}, it depends on the performance property; for \performance{cpu}, \performance{fps} and \performance{time}, the effect of influential options are stable for all inputs, while for the \performance{bitrate} and the \performance{size}, we can draw the conclusions previously presented. 

\begin{tcolorbox}[boxsep=-2pt]
\textbf{RQ$_{\mathbf{2}}$ - \rqfeature}
Two lessons learned :
(1) the importance of software options change with input data, implying that an option can be influential only for few input data, but not for the rest of the inputs;
(2) the effect of software options on performance properties vary with input data. An option can have a positive influence for an input and at the same time a negative influence for another input. 
As a result, tuning the options of a software system should depend on its processed inputs. 
\end{tcolorbox}

%%%%%%%%%%%%%%%%%%%%%%%%%%%%%%%%%%%%%% RQ3 %%%%%%%%%%%%%%%%%%%%%%%%%%%%%%%%%%%%%%%%%%%%%%%%%%%

\subsection{Impact of Inputs on Performance ($RQ_{3}$)}
\label{sub:results:impact}

\begin{figure}[ht]
    % \vspace*{-0.3cm}
    \centering
    \includegraphics[width=\linewidth]{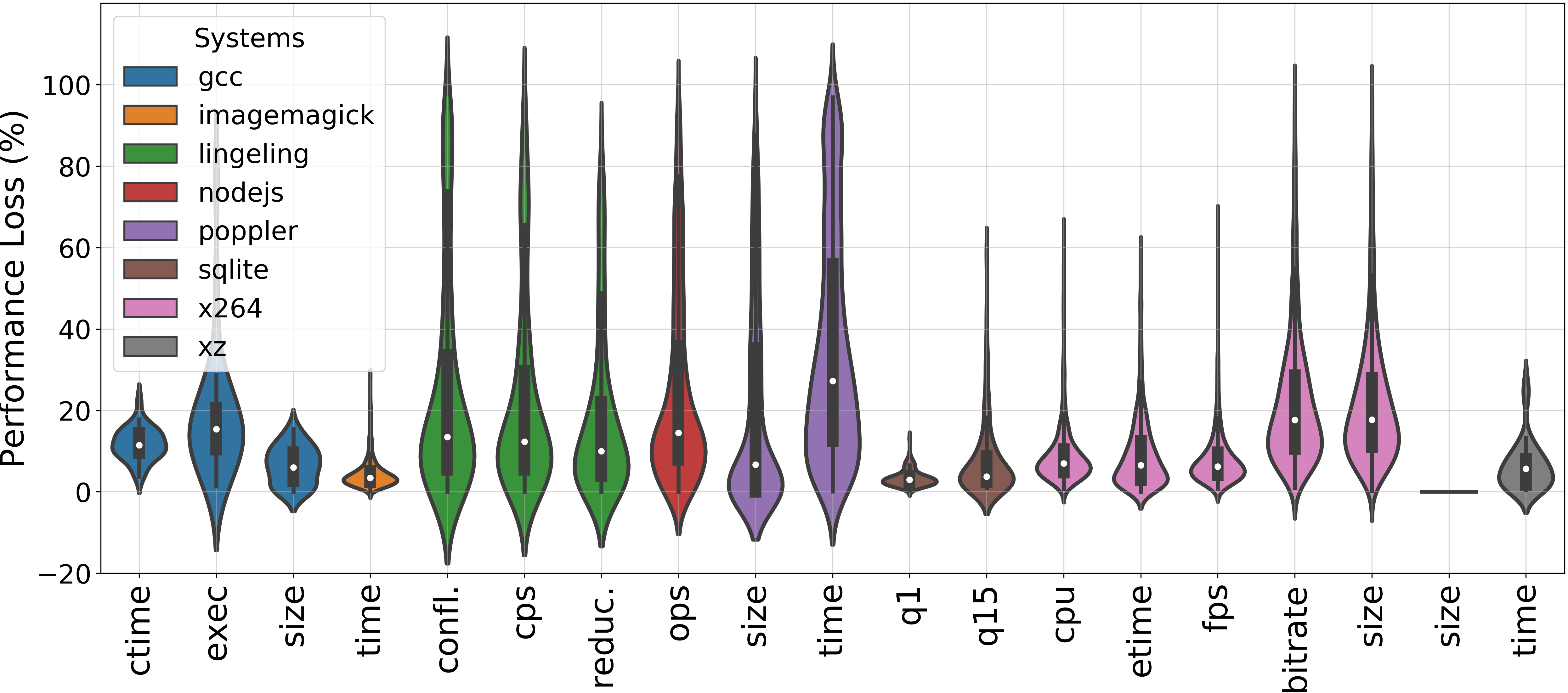}
    % \vspace*{-0.5cm}
    \caption{Performance loss (\%, y-axis) when ignoring input sensitivity per system and performance property (x-axis)}
    \label{fig:violin_plot}
    % \vspace*{-0.3cm}
\end{figure}

This section presents the evaluation of $RQ_{3}$\footnote{Detailed $RQ_{3}$ results for other performance properties are available at: \scriptsize{\bluerl{https://github.com/llesoil/input_sensitivity/tree/master/results/RQS/RQ3/RQ3.md}}} \wrt the protocol of \Cref{sub:protocol:impact}. 
\Cref{fig:violin_plot} presents the loss of performance (y-axis, in \%) due to input sensitivity for the different software systems and their performance properties.

%input{figure/RQ3-table_ratios}
%input{figure/RQ3-violin_plot}

\textbf{Key result.} 
The average performance ratio across all the software systems is \num{1.38}: we can expect an average drop of \pc{38} in terms of performance when ignoring the input sensitivity\footnote{To compute this result, we removed \soft{SQLite} biasing the results with its \num{15} performance properties}. 

\textbf{Meta-analysis.} 
For software systems whose performance are stable across inputs (\soft{gcc}, \soft{imagemagick} and \soft{xz}), there are few differences between inputs.
For instance, for the output \performance{size} of \soft{xz}, there is no variation between scenarios $S_{1}$ (\ie using the best configuration) and $S_{2}$ (\ie reusing a the best configuration of a given input for another input): all performance ratios (\ie performance $S_{1}$ over performance $S_{2}$) are equals to \num{1} whatever the input.

For input-sensitive software systems (\soft{lingeling}, \soft{nodeJS}, \soft{SQLite} and \soft{poppler}), changing the configuration can lead to a negligible change in a few cases.
For instance, for the time to answer the first query \performance{q1} with \soft{SQLite}, the median is \num{1.03}; in this case, \soft{SQLite} is sensitive to inputs, but its variations of performance -less than \pc{4}- do not justify the complexity of tuning the software. 
But it can also be a huge change; for \soft{lingeling} and solved \performance{conflicts}, the $95^{th}$ percentile ratio is equal to \num{8.05} \ie a factor of \num{8} between $S_{1}$ and $S_{2}$.
It goes up to a ratio of \num{10.11} for \soft{poppler}'s extraction \performance{time}: there exists an input pdf for which extracting its images is ten times slower when reusing a configuration compared to the fastest. 

In between, \soft{x264} is a complex case. 
For its low input-sensitive performance (\eg \performance{cpu} and \performance{etime}), it moderately impacts the performance when reusing a configuration from one input to another - average ratios at \resp \num{1.42} and \num{1.43}. 
In this case, the rankings of performance do not change a lot with inputs, but a small ranking change does make the difference in terms of performance.

On the contrary, for the input-sensitive performance (\eg the \performance{bitrate}), there are few variations of performance: we can lose $1-\frac{1}{1.11} \simeq $ \pc{9} of \performance{bitrate} in average. 
In this case, it is up to the compression experts to decide; if losing up to $1-\frac{1}{1.32} \simeq $ \pc{24} of \performance{bitrate} is acceptable, then we can ignore input sensitivity. 
Otherwise, we should consider tuning \soft{x264} for its input video. 

\vspace{-0.1cm}

\begin{tcolorbox}[boxsep=-2pt]
\textbf{RQ$_{\mathbf{3}}$ - \rqimpact}
In average, randomly reusing configurations across inputs leads to a performance drop of \pc{38}, which suggests we cannot ignore input sensitivity. 
On the good side, performance can be multiplied up to a ratio of \num{10} if we tune other systems for their input data. 
\end{tcolorbox}

%%%%%%%%%%%%%%%%%%%%%%%%%%%%%%%%%%%%%% RQ4 %%%%%%%%%%%%%%%%%%%%%%%%%%%%%%%%%%%%%%%%%%%%%%%%%%%

\subsection{Groups of Inputs ($RQ_{4}$)} 
\label{sub:results:groups}

\begin{figure}[h]
    \centering
    \includegraphics[width=\columnwidth]{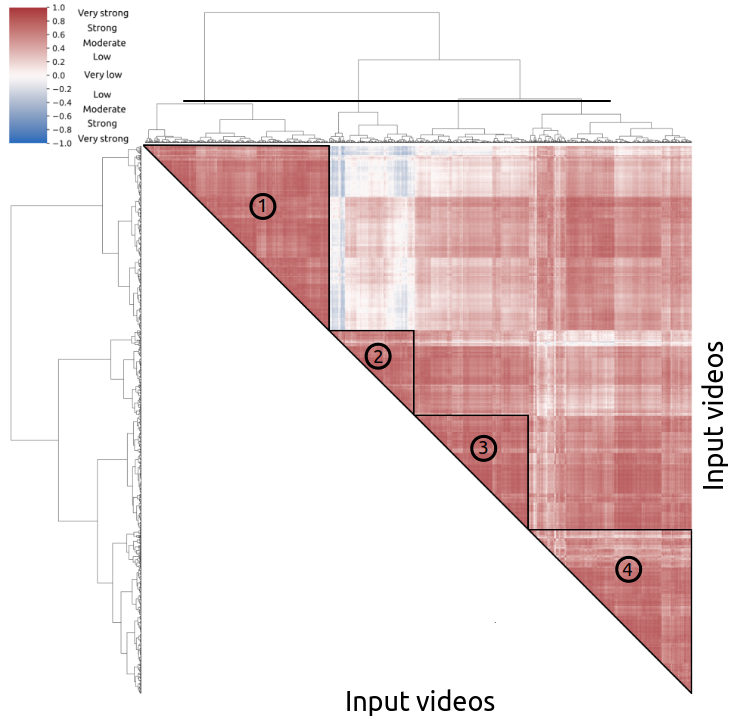}
    \caption{Performance groups of input videos - \soft{x264}, \performance{bitrate}}
    \label{fig:corr_group}
    % \vspace*{-0.3cm}
\end{figure}

We illustrate the results of this section using the \performance{bitrate} of \soft{x264} when encoding input videos\footnote{The results for the rest of software systems can be consulted in the companion repository at \url{https://github.com/llesoil/input_sensitivity/blob/master/results/RQS/RQ4/groups.ipynb}}. 
In \Cref{fig:corr_group}, we first compute the correlations between performance of all input videos, as in $RQ_{1}$. Then, we perform hierarchical clustering on \soft{x264} measurements to gather inputs having similar \performance{bitrate} distributions and visually group correlated videos together. 
The resulting groups are delimited and numbered directly in the figure. For instance, the group \raisebox{.5pt}{\textcircled{\raisebox{-.9pt} {1}}} is located in the top-left part of the correlogram  by the triangle \raisebox{.5pt}{\textcircled{\raisebox{-.9pt} {1}}}). 

% \vspace*{-0.3cm}

\begin{table}[h]
\renewcommand{\arraystretch}{0.9}
\setlength{\tabcolsep}{1pt}
\centering
\caption{Performance groups of input videos - \soft{x264}, \performance{bitrate}}
\label{tab:input_group}
% \vspace*{-0.3cm}
% \hspace*{-0.3cm}
\begin{tabular}{|c|c|c|c|c|c|}
\hline
\multicolumn{2}{|c|}{\scriptsizeb {\textbf{Group}}}
 & \scriptsizeb{ \textbf{\raisebox{.5pt}{\textcircled{\raisebox{-.9pt} {1}}} Action }}
 & \scriptsizeb{ \textbf{\raisebox{.5pt}{\textcircled{\raisebox{-.9pt} {2}}} Big}}
 & \scriptsizeb{ \textbf{\raisebox{.5pt}{\textcircled{\raisebox{-.9pt} {3}}} Still image}}
 & \scriptsizeb{ \textbf{\raisebox{.5pt}{\textcircled{\raisebox{-.9pt} {4}}} Standard}}
 \tabularnewline \hline
\multicolumn{2}{|c|}{\scriptsizeb{\# Inputs}}
 & \scriptsizeb{ 470}
 & \scriptsizeb{ 219}
 & \scriptsizeb{ 292}
 & \scriptsizeb{ 416}
 \tabularnewline \hline
\multicolumn{2}{|c|}{\multirow{3}{*}{\scriptsizeb{ Input Properties}}}
 & \scriptsizeb{ Spatial ++ }
 & \scriptsizeb{ Spatial \mbox{–}\mbox{–}}
 & \scriptsizeb{ Spatial \mbox{–}\mbox{–}}
 & \scriptsizeb{ Width -}
 \\
\multicolumn{2}{|c|}{}
 & \scriptsizeb{ Chunk ++ }
 & \scriptsizeb{ Temporal ++}
 & \scriptsizeb{ Temporal \mbox{–}\mbox{–}}
 & \scriptsizeb{ Height -}
 \\
\multicolumn{2}{|c|}{}
 & 
 & \scriptsizeb{ Width ++}
 & \scriptsizeb{ Chunk \mbox{–}\mbox{–}}
 & \scriptsizeb{ Temporal -}
 \tabularnewline \hline
\multicolumn{2}{|c|}{\multirow{2}{*}{\scriptsizeb{ Main Category}}}
 & \scriptsizeb{ Sports}
 & \scriptsizeb{ HDR}
 & \scriptsizeb{ Lecture}
 & \scriptsizeb{ Music}
 \tabularnewline
\multicolumn{2}{|c|}{}
 & \scriptsizeb{ News}
 &  
 & \scriptsizeb{ HowTo}
 & \scriptsizeb{ Vertical}
 \tabularnewline \hline
 \multicolumn{2}{|c|}{\scriptsizeb{ Avg Correlation}}
 & \multirow{2}{*}{\scriptsizeb{ \textbf{0.82} $\pm$ 0.11}}
 & \multirow{2}{*}{\scriptsizeb{ \textbf{0.79} $\pm$ 0.14}}
 & \multirow{2}{*}{\scriptsizeb{ \textbf{0.85} $\pm$ 0.09}}
 & \multirow{2}{*}{\scriptsizeb{ \textbf{0.74} $\pm$ 0.17}}
 \tabularnewline
\multicolumn{2}{|c|}{\scriptsizeb{ (\eg \cref{fig:spearmansize})}}
 &  
 & 
 & 
 & 
 \tabularnewline \hline
\scriptsizeb{ Imp.} & \scriptsizeb{\mbox{–}\mbox{–}\feature{mbtree}}
 & \scriptsizeb{ \textbf{0.09} $\pm$ 0.09}
 & \scriptsizeb{ \textbf{0.47} $\pm$ 0.2}
 & \scriptsizeb{ \textbf{0.34} $\pm$ 0.22}
 & \scriptsizeb{ \textbf{0.05} $\pm$ 0.07}
 \tabularnewline \cline{2-6}
\scriptsizeb{ (\cref{fig:featurerfboxplot-kbs})} & \scriptsizeb{\mbox{–}\mbox{–}\feature{aq-mode}}
 & \scriptsizeb{ \textbf{0.27} $\pm$ 0.19}
 & \scriptsizeb{ \textbf{0.13} $\pm$ 0.13}
 & \scriptsizeb{ \textbf{0.04} $\pm$ 0.07}
 & \scriptsizeb{ \textbf{0.15} $\pm$ 0.18}
 \tabularnewline \hline
  \scriptsizeb{ Effect}
 & \scriptsizeb{\mbox{–}\mbox{–}\feature{mbtree}}
 & \scriptsizeb{ \textbf{0.33} $\pm$ 0.19}
 & \scriptsizeb{ \textbf{-0.68} $\pm$ 0.18}
 & \scriptsizeb{ \textbf{-0.42} $\pm$ 0.15}
 & \scriptsizeb{ \textbf{-0.11} $\pm$ 0.15}
 \tabularnewline \cline{2-6}
\scriptsizeb{ (\cref{fig:featurelinboxplot-kbs})} &  \scriptsizeb{\mbox{–}\mbox{–}\feature{aq-mode}}
 & \scriptsizeb{ \textbf{-0.5} $\pm$ 0.14}
 & \scriptsizeb{ \textbf{0.36} $\pm$ 0.21}
 & \scriptsizeb{ \textbf{-0.14} $\pm$ 0.14}
 & \scriptsizeb{ \textbf{-0.29} $\pm$ 0.18}
 \tabularnewline \hline
\end{tabular}
% \vspace*{-0.4cm}
\end{table}

\textbf{Group description.}
In total, we isolate four groups of input videos. These groups are presented and described in \Cref{tab:input_group}:
\begin{itemize}[topsep=0pt,itemsep=-1ex,partopsep=1ex,parsep=1ex, leftmargin=0.3cm]
    \item Group \raisebox{.5pt}{\textcircled{\raisebox{-.9pt} {1}}} is mostly composed of moving or action videos, often picked in the sports or news categories and with high spatial and chunk complexities; 
    \item Group \raisebox{.5pt}{\textcircled{\raisebox{-.9pt} {2}}} gathers large input videos, with big resolution videos, taken for instance in the High Dynamic Range category. They typically have a low spatial complexity and a high temporal complexity;
    \item Group \raisebox{.5pt}{\textcircled{\raisebox{-.9pt} {3}}} is composed of "still image" videos \ie input videos with few changes of background, with low temporal and chunk complexities. A typical example of this kind of video would be a course with a fixed board, chosen in the Lecture or in the HowTo category;
    \item Group \raisebox{.5pt}{\textcircled{\raisebox{-.9pt} {4}}} is a group of average videos with average properties values and various contents. 
\end{itemize}
\bigskip

\textbf{Revisiting $RQ_{1}$.}
% groupes d'inputs corrélés en intra
In a group of inputs, performance distributions of inputs are highly correlated with each other - positively, strong or very strong. 
The input videos of the same group have similar \performance{bitrate} rankings; their performance react the same way to the same configurations of \soft{x264}. 
% mais pas forcément en inter 
However, the group \raisebox{.5pt}{\textcircled{\raisebox{-.9pt} {1}}} is uncorrelated (very low, low) or negatively correlated (moderate, strong and very strong) with the group \raisebox{.5pt}{\textcircled{\raisebox{-.9pt} {2}}} - see the intersection area between triangles \raisebox{.5pt}{\textcircled{\raisebox{-.9pt} {1}}} and \raisebox{.5pt}{\textcircled{\raisebox{-.9pt} {2}}}. 
In this case, a single configuration of \soft{x264} working for the group \raisebox{.5pt}{\textcircled{\raisebox{-.9pt} {1}}} should not be reused directly on a video of the group \raisebox{.5pt}{\textcircled{\raisebox{-.9pt} {2}}}. 
So, these groups are capturing the difference of performance between inputs; once in a group, input sensitivity does not represent a problem anymore.

\textbf{Revisiting $RQ_{2}$.} 
Within a group, the effect and importance of options are stable and the inputs all react the same way to the same options, while they differ between the different groups.
For instance, for the group \raisebox{.5pt}{\textcircled{\raisebox{-.9pt} {1}}}, \feature{aq-mode} is influential (Imp = $0.27$), while it is not for the group \raisebox{.5pt}{\textcircled{\raisebox{-.9pt} {3}}} (Imp = $0.04$).
Likewise, the effects of \feature{mbtree} vary with the group of inputs; for the group \raisebox{.5pt}{\textcircled{\raisebox{-.9pt} {1}}}, activating \feature{mbtree} always increases the \performance{bitrate} (Effect = \textbf{+}$0.33$), while for the groups \raisebox{.5pt}{\textcircled{\raisebox{-.9pt} {2}}}, \raisebox{.5pt}{\textcircled{\raisebox{-.9pt} {3}}} and \raisebox{.5pt}{\textcircled{\raisebox{-.9pt} {4}}}, it diminishes the \performance{bitrate} (Effects = \textbf{-}$0.68$, \textbf{-}$0.42$, \textbf{-}$0.11$ respectively).
Under these circumstances, configuring the software system once per group of inputs is probably a reasonable solution for tackling input sensitivity. 

\textbf{Revisiting $RQ_{3}$.} 
For the \performance{bitrate} of \soft{x264}, reusing a configuration from a source input to a target input generate a lower performance drop if the source and the target inputs are selected in the same group (\eg \pc{13} for group 2) compared to a random selection (\pc{34} in general). 
If we are able to find the best configuration for one input video in a group, this configuration will be good-enough for the rest of the inputs in this group.

\textbf{Meta-analysis.}
For the other input-sensitive systems, results tend to show similar results as for \soft{x264} and the \performance{bitrate}. For instance, with \soft{poppler}, grouping tend to gather inputs with the same influence and effect of options; for the \performance{size}, the importance of format is influential in groups 2 and 4 but not in groups 1 and 3; for the execution \performance{time}, in groups 1 and 2, \mbox{-}jp2 has a positive effect overall while a negative effect for groups 3 and 4. When the groups discriminate inputs with different effects of options, $RQ_{3}$ results tend to be more impressive \eg the average performance loss of \pc{49} vanishes when grouping the inputs (in the four groups, \pc{1}-\pc{4}-\pc{12}-\pc{2} when reusing a configuration inside the group). 
The same applies for \soft{Nodejs} : \mbox{-}\mbox{-}jitless is influential for groups 2, 3 and 4 but not for group 1. In $RQ_{3}$ results, the average performance drop of \pc{44} becomes \pc{7.4}-\pc{6.7}-\pc{13.7}-\pc{9.0} in the four groups.
For non input-sensitive systems, we do not observe such difference between the groups. Grouping seems to be ineffective; the same effect of options are observed; it does not change the performance loss, already low \eg for the execution \soft{time} of \soft{imagemagick}, \pc{6} in general and \pc{1}-\pc{1}-\pc{2}-\pc{6} in the groups.

\textbf{Classify inputs into groups.}
In short, grouping together inputs seems a right approach to reduce input sensitivity. 
However there is now a problem: we need to map a given input into a group \emph{a priori}, without having access to all measurements. 
Since these four groups are consistent and share common properties, one domain expert or one machine learning model could classify these inputs \textit{a priori} into a group without measuring their performance just by looking at the properties of the inputs.

\textbf{Benefits for benchmarking.} 
These groups allow to increase the representativeness of profiles of inputs used to test software systems, while greatly lowering the number of inputs of this set.
In the companion repository, we operate on previous results to create a short but representative set of input videos dedicated to the benchmarking of \soft{x264}: we reduce the dataset, initially composed of \num{1397} input videos \cite{Wang2019}, to a subset of \num{8} videos, selecting \num{2} cheap videos in each group of performance\footnote{See the resulting benchmark and its construction at: \scriptsize{\bluerl{https://github.com/llesoil/input_sensitivity/tree/master/results/RQS/RQ4/x264_bitrate.md}}}. 

\begin{tcolorbox}[boxsep=-2pt]
\textbf{RQ$_{\mathbf{4}}$ - \rqgroup}
Grouping inputs together is beneficial to apply in input-sensitive systems: the performance distributions, the influence of options, and the effect of options are alike between inputs of the same group. To classify inputs in one of those groups, the characteristics of inputs can be used without measuring any configuration. These groups can also be derived to create short but representative sets of inputs designed to benchmark software systems. For other systems, it does not bring any significant improvements.
\end{tcolorbox}

\section{Sensitivity to Inputs in Research}
\label{sec:significance}

%input{figure/RQ5-table_paper}
\begin{table*}[htp]
\centering

\setlength{\tabcolsep}{2.5pt}

\caption{\label{tab:papers}Input sensitivity in research. \textbf{\textit{Q-A.}} \pqone~\textbf{\textit{Q-B.}} \pqtwo~\textbf{\textit{Q-C.}} \pqthree~\textbf{\textit{Q-D.}} \pqfour ~ \textcolor{blue}{\href{https://github.com/llesoil/input_sensitivity/tree/master/results/RQS/RQ6/RQ6.md}{Justifications in the companion repository}}.}
\vspace*{-0.2cm}

\renewcommand{\arraystretch}{0.58}

% \hspace*{-3cm}
\begin{tabular}{|>{\centering}p{3mm}|>{\centering}p{3cm}|>{\centering}p{1.5cm}|>{\centering}p{6mm}|>{\centering}p{8.5cm}|>{\centering}p{6mm}|>{\centering}p{6mm}|>{\centering}p{6mm}|>{\centering}p{6mm}|}
\hline

\scriptsize \cellcolor[HTML]{e8e8e8}{\textbf{ID}}
 & \scriptsize \cellcolor[HTML]{e8e8e8}{\textbf{Authors}}
 & \scriptsize \cellcolor[HTML]{e8e8e8}{\textbf{Conference}}
 & \scriptsize \cellcolor[HTML]{e8e8e8}{\textbf{Year}}
 & \scriptsize \cellcolor[HTML]{e8e8e8}{\textbf{Title}}
 & \scriptsize \cellcolor[HTML]{e8e8e8}{\textbf{Q-A}}
%: Quels sont les systèmes configurables considérés? (et ensuite on regarde: Sont-ils sujets à l'input sensitivity?) "
% & Système(s) configurable(s)	
 & \scriptsize \cellcolor[HTML]{e8e8e8}{\textbf{Q-B}}
%. Est-ce que les modèles de performances sont appliqués sur plusieurs inputs? Est-ce que plusieurs inputs sont considérés? (aka est-ce que le papier considère plusieurs inputs dans l'évaluation et/ou l'approche?)"	
 & \scriptsize \cellcolor[HTML]{e8e8e8}{\textbf{Q-C}}
%Est-ce que le problème de l'input sensitivity est mentionné (eg en threat)?
 & \scriptsize \cellcolor[HTML]{e8e8e8}{\textbf{Q-D}} \tabularnewline 
% Est-ce le problème de l'input sensitivité est résolu? aka Est-ce qu'il y a un modèle de perf qui marche pour tous les inputs?
\hline

\scriptsize 1 & \scriptsize Guo \etal~\cite{Guo2017} & \scriptsize ESE & \scriptsize 2017 & \scriptsize Data-efficient performance learning for configurable systems & \scriptsize X & \scriptsize  & \scriptsize  & \scriptsize 	\tabularnewline \hline 

\scriptsize 2 & \scriptsize Jamshidi \etal~\cite{JVKSK:SEAMS17} & \scriptsize SEAMS & \scriptsize 2017 & \scriptsize Transfer learning for improving model predictions [...] & \scriptsize X & \scriptsize X & \scriptsize X & \scriptsize 	\tabularnewline \hline 

\scriptsize 3 & \scriptsize Jamshidi \etal~\cite{jamshidi2017b} & \scriptsize ASE & \scriptsize 2017 & \scriptsize Transfer learning for performance modeling of configurable [...] & \scriptsize X & \scriptsize X & \scriptsize X & \scriptsize X	\tabularnewline \hline 

\scriptsize 4 & \scriptsize Oh \etal~\cite{10.1145/3106237.3106273} & \scriptsize ESEC/FSE & \scriptsize 2017 & \scriptsize Finding near-optimal configurations in product lines by [...] & \scriptsize X & \scriptsize  & \scriptsize  & \scriptsize 	\tabularnewline \hline 

\scriptsize 5 & \scriptsize Kolesnikov \etal~\cite{Kolesnikov2018} & \scriptsize SoSyM & \scriptsize 2018 & \scriptsize Tradeoffs in modeling performance of highly configurable [...] & \scriptsize X & \scriptsize  & \scriptsize  & \scriptsize 	\tabularnewline \hline 

\scriptsize 6 & \scriptsize Nair \etal~\cite{nair2017} & \scriptsize ESEC/FSE & \scriptsize 2017 & \scriptsize Using bad learners to find good configurations & \scriptsize X & \scriptsize X & \scriptsize  & \scriptsize 	\tabularnewline \hline 

\scriptsize 7 & \scriptsize Nair \etal~\cite{flash_find_config} & \scriptsize TSE & \scriptsize 2018 & \scriptsize Finding Faster Configurations using FLASH & \scriptsize X & \scriptsize X & \scriptsize X & \scriptsize 	\tabularnewline \hline 

\scriptsize 8 & \scriptsize Murwantara \etal~\cite{10.1145/2684200.2684314} & \scriptsize iiWAS & \scriptsize 2014 & \scriptsize Measuring Energy Consumption for Web Service Product [...] & \scriptsize X & \scriptsize X & \scriptsize X & \scriptsize X	\tabularnewline \hline 

\scriptsize 9 & \scriptsize Temple \etal~\cite{Temple2016} & \scriptsize SPLC & \scriptsize 2016 & \scriptsize Using Machine Learning to Infer Constraints for Product Lines & \scriptsize  & \scriptsize \cellcolor[HTML]{e8e8e8}{} & \scriptsize \cellcolor[HTML]{e8e8e8}{} & \scriptsize \cellcolor[HTML]{e8e8e8}{}	\tabularnewline \hline 

\scriptsize 10 & \scriptsize Temple \etal~\cite{IEEEcontextTemple} & \scriptsize IEEE Soft. & \scriptsize 2017 & \scriptsize Learning Contextual-Variability Models & \scriptsize X & \scriptsize  & \scriptsize X & \scriptsize 	\tabularnewline \hline 

\scriptsize 11 & \scriptsize Valov \etal~\cite{DBLP:conf/wosp/ValovPGFC17} & \scriptsize ICPE & \scriptsize 2017 & \scriptsize Transferring performance prediction models across different [...] & \scriptsize X & \scriptsize  & \scriptsize X & \scriptsize X \tabularnewline \hline 

\scriptsize 12 & \scriptsize Weckesser \etal~\cite{10.1145/3233027.3233030} & \scriptsize SPLC & \scriptsize 2018 & \scriptsize Optimal reconfiguration of dynamic software product [...] & \scriptsize  & \scriptsize \cellcolor[HTML]{e8e8e8}{} & \scriptsize \cellcolor[HTML]{e8e8e8}{} & \scriptsize \cellcolor[HTML]{e8e8e8}{}	\tabularnewline \hline 

\scriptsize 13 & \scriptsize Acher \etal~\cite{10.1145/3168365.3168372} & \scriptsize VaMoS & \scriptsize 2018 & \scriptsize VaryLATEX: Learning Paper Variants That Meet Constraints & \scriptsize X & \scriptsize X & \scriptsize  & \scriptsize 	\tabularnewline \hline 

\scriptsize 14 & \scriptsize Sarkar \etal~\cite{guo2015} & \scriptsize ASE & \scriptsize 2015 & \scriptsize Cost-Efficient Sampling for Performance Prediction of [...] & \scriptsize X & \scriptsize  & \scriptsize  & \scriptsize 	\tabularnewline \hline 

\scriptsize 15 & \scriptsize Temple \etal~\cite{temple2018adversarial} & \scriptsize Report & \scriptsize 2018 & \scriptsize Towards Adversarial Configurations for Software Product Lines & \scriptsize  & \scriptsize \cellcolor[HTML]{e8e8e8}{} & \scriptsize \cellcolor[HTML]{e8e8e8}{} & \scriptsize \cellcolor[HTML]{e8e8e8}{}	\tabularnewline \hline 

\scriptsize 16 & \scriptsize Nair \etal~\cite{10.1007/s10515-017-0225-2} & \scriptsize ASE & \scriptsize 2018 & \scriptsize Faster Discovery of Faster System Configurations with [...] & \scriptsize X & \scriptsize  & \scriptsize  & \scriptsize 	\tabularnewline \hline 

\scriptsize 17 & \scriptsize Siegmund \etal~\cite{SGKA:ESECFSE15} & \scriptsize ESEC/FSE & \scriptsize 2015 & \scriptsize Performance-Influence Models for Highly Configurable Systems & \scriptsize X & \scriptsize  & \scriptsize  & \scriptsize 	\tabularnewline \hline 

\scriptsize 18 & \scriptsize Valov \etal~\cite{DBLP:conf/splc/ValovGC15} & \scriptsize SPLC & \scriptsize 2015 & \scriptsize Empirical comparison of regression methods for [...] & \scriptsize X & \scriptsize  & \scriptsize  & \scriptsize 	\tabularnewline \hline 

\scriptsize 19 & \scriptsize Zhang \etal~\cite{DBLP:conf/kbse/ZhangGBC15} & \scriptsize ASE & \scriptsize 2015 & \scriptsize Performance Prediction of Configurable Software Systems [...] & \scriptsize X & \scriptsize  & \scriptsize X & \scriptsize 	\tabularnewline \hline 

\scriptsize 20 & \scriptsize Kolesnikov \etal~\cite{kolesnikov2018relation} & \scriptsize ESE & \scriptsize 2019 & \scriptsize On the relation of control-flow and performance feature [...] & \scriptsize X & \scriptsize  & \scriptsize  & \scriptsize 	\tabularnewline \hline 

\scriptsize 21 & \scriptsize Couto \etal~\cite{10.1145/3106195.3106214} & \scriptsize SPLC & \scriptsize 2017 & \scriptsize Products go Green: Worst-Case Energy Consumption [...] & \scriptsize X & \scriptsize  & \scriptsize X & \scriptsize 	\tabularnewline \hline 

\scriptsize 22 & \scriptsize Van Aken \etal~\cite{10.1145/3035918.3064029} & \scriptsize SIGMOD & \scriptsize 2017 & \scriptsize Automatic Database Management System Tuning Through [...] & \scriptsize X & \scriptsize X & \scriptsize X & \scriptsize X \tabularnewline \hline 

\scriptsize 23 & \scriptsize Kaltenecker \etal~\cite{10.1109/ICSE.2019.00112} & \scriptsize ICSE & \scriptsize 2019 & \scriptsize Distance-based sampling of software configuration spaces & \scriptsize X & \scriptsize  & \scriptsize  & \scriptsize 	\tabularnewline \hline

\scriptsize 24 & \scriptsize Jamshidi \etal~\cite{jamshidi2018} & \scriptsize ESEC/FSE & \scriptsize 2018 & \scriptsize Learning to sample: exploiting similarities across [...] & \scriptsize X & \scriptsize X & \scriptsize X & \scriptsize X	\tabularnewline \hline

\scriptsize 25 & \scriptsize Jamshidi \etal~\cite{DBLP:journals/corr/JamshidiC16} & \scriptsize MASCOTS & \scriptsize 2016 & \scriptsize An Uncertainty-Aware Approach to Optimal Configuration of [...] & \scriptsize X & \scriptsize X & \scriptsize X & \scriptsize 	\tabularnewline \hline

\scriptsize 26 & \scriptsize Lillacka \etal~\cite{lillack2013improved} & \scriptsize Soft. Eng. & \scriptsize 2013 & \scriptsize Improved prediction of non-functional properties in Software [...] & \scriptsize X & \scriptsize X & \scriptsize X & \scriptsize X	\tabularnewline \hline

\scriptsize 27 & \scriptsize Zuluaga \etal~\cite{JMLR:v17:15-047} & \scriptsize JMLR & \scriptsize 2016 & \scriptsize $\varepsilon$-pal: an active learning approach [...] & \scriptsize X & \scriptsize X & \scriptsize  & \scriptsize 	\tabularnewline \hline

\scriptsize 28 & \scriptsize Amand \etal~\cite{10.1145/3302333.3302338} & \scriptsize VaMoS & \scriptsize 2019 & \scriptsize Towards Learning-Aided Configuration in 3D Printing [...] & \scriptsize X & \scriptsize X & \scriptsize X & \scriptsize \tabularnewline \hline

\scriptsize 29 & \scriptsize Alipourfard \etal~\cite{10.5555/3154630.3154669} & \scriptsize NSDI & \scriptsize 2017 & \scriptsize Cherrypick: Adaptively unearthing the best cloud [...] & \scriptsize X & \scriptsize X & \scriptsize X & \scriptsize 	\tabularnewline \hline 

\scriptsize 30 & \scriptsize Saleem \etal~\cite{6981951} & \scriptsize TSC & \scriptsize 2015 & \scriptsize Personalized Decision-Strategy based Web Service Selection [...] & \scriptsize X & \scriptsize X & \scriptsize  & \scriptsize 	\tabularnewline \hline 

\scriptsize 31 & \scriptsize Zhang \etal~\cite{10.1145/2934466.2934469} & \scriptsize SPLC & \scriptsize 2016 & \scriptsize A mathematical model of performance-relevant [...] & \scriptsize X & \scriptsize  & \scriptsize  & \scriptsize 	\tabularnewline \hline

\scriptsize 32 & \scriptsize Ghamizi \etal~\cite{10.1145/3336294.3336306} & \scriptsize SPLC & \scriptsize 2019 & \scriptsize Automated Search for Configurations of Deep Neural [...] & \scriptsize X & \scriptsize X & \scriptsize X & \scriptsize 	\tabularnewline \hline

\scriptsize 33 & \scriptsize Grebhahn \etal~\cite{Grebhahn2017PerformanceinfluenceMO} & \scriptsize CPE & \scriptsize 2017 & \scriptsize Performance-influence models of multigrid methods [...] & \scriptsize  & \scriptsize \cellcolor[HTML]{e8e8e8}{} & \scriptsize \cellcolor[HTML]{e8e8e8}{} & \scriptsize \cellcolor[HTML]{e8e8e8}{}	\tabularnewline \hline

\scriptsize 34 & \scriptsize Bao \etal~\cite{10.1145/3238147.3238175} & \scriptsize ASE & \scriptsize 2018 & \scriptsize AutoConfig: Automatic Configuration Tuning for Distributed [...] & \scriptsize X & \scriptsize X & \scriptsize  & \scriptsize 	\tabularnewline \hline 

\scriptsize 35 & \scriptsize Guo \etal~\cite{guo2013} & \scriptsize ASE & \scriptsize 2013 & \scriptsize Variability-aware performance prediction: A statistical [...] & \scriptsize X & \scriptsize  & \scriptsize  & \scriptsize 	\tabularnewline \hline

\scriptsize 36 & \scriptsize Švogor \etal~\cite{SVOGOR201930} & \scriptsize IST & \scriptsize 2019 & \scriptsize An extensible framework for software configuration optim[...] & \scriptsize X & \scriptsize X & \scriptsize  & \scriptsize 	\tabularnewline \hline 

\scriptsize 37 & \scriptsize El Afia \etal~\cite{8284699} & \scriptsize CloudTech & \scriptsize 2018 & \scriptsize Performance prediction using support vector machine for the [...] & \scriptsize X & \scriptsize X & \scriptsize  & \scriptsize 	\tabularnewline \hline

\scriptsize 38 & \scriptsize Ding \etal~\cite{ding2015} & \scriptsize PLDI & \scriptsize 2015 & \scriptsize Autotuning algorithmic choice for input sensitivity & \scriptsize X & \scriptsize X & \scriptsize X & \scriptsize X	\tabularnewline \hline 

\scriptsize 39 & \scriptsize Duarte \etal~\cite{10.1145/3194133.3194138} & \scriptsize SEAMS & \scriptsize 2018 & \scriptsize Learning Non-Deterministic Impact Models for Adaptation & \scriptsize X & \scriptsize X & \scriptsize X & \scriptsize X	\tabularnewline \hline

\scriptsize 40 & \scriptsize Thornton \etal~\cite{thornton2013} & \scriptsize KDD & \scriptsize 2013 & \scriptsize Auto-WEKA: Combined selection and hyperparameter [...] & \scriptsize X & \scriptsize X & \scriptsize X & \scriptsize 	\tabularnewline \hline 

\scriptsize 41 & \scriptsize Siegmund \etal~\cite{siegmund2012predicting} & \scriptsize ICSE & \scriptsize 2012 & \scriptsize Predicting performance via automated feature-inter[...] & \scriptsize X & \scriptsize X & \scriptsize X & \scriptsize 	\tabularnewline \hline 

\scriptsize 42 & \scriptsize Siegmund \etal~\cite{10.1007/s11219-011-9152-9} & \scriptsize SQJ & \scriptsize 2012 & \scriptsize SPL Conqueror: Toward optimization of non-functional [...] & \scriptsize X & \scriptsize X & \scriptsize  & \scriptsize 	\tabularnewline \hline 

\scriptsize 43 & \scriptsize Westermann \etal~\cite{10.1145/2351676.2351703} & \scriptsize ASE & \scriptsize 2012 & \scriptsize Automated inference of goal-oriented performance prediction [...] & \scriptsize X & \scriptsize X & \scriptsize  & \scriptsize 	\tabularnewline \hline 

\scriptsize 44 & \scriptsize Velez \etal~\cite{velez2021whitebox} & \scriptsize ICSE & \scriptsize 2021 & \scriptsize White-Box Analysis over Machine Learning: Modeling [...] & \scriptsize X & \scriptsize X & \scriptsize  & \scriptsize 	\tabularnewline \hline 

\scriptsize 45 & \scriptsize Pereira \etal~\cite{10.1145/3358960.3379137} & \scriptsize ICPE & \scriptsize 2020 & \scriptsize Sampling Effect on Performance Prediction of Configurable [...] & \scriptsize X & \scriptsize X & \scriptsize X & \scriptsize 	\tabularnewline \hline 

\scriptsize 46 & \scriptsize Shu \etal~\cite{10.1145/3382494.3410677} & \scriptsize ESEM & \scriptsize 2020 & \scriptsize Perf-AL: Performance prediction for configurable software [...] & \scriptsize X & \scriptsize  & \scriptsize  & \scriptsize 	\tabularnewline \hline 

\scriptsize 47 & \scriptsize Dorn \etal~\cite{10.1145/3324884.3416620} & \scriptsize ASE & \scriptsize 2020 & \scriptsize Mastering Uncertainty in Performance Estimations of [...] & \scriptsize X & \scriptsize  & \scriptsize  & \scriptsize	\tabularnewline \hline 

\scriptsize 48 & \scriptsize Kaltenecker \etal\cite{DBLP:journals/software/KalteneckerGSA20} & \scriptsize IEEE Soft. & \scriptsize 2020 & \scriptsize The Interplay of Sampling and Machine Learning for Software [...] & \scriptsize X & \scriptsize  & \scriptsize  & \scriptsize 	\tabularnewline \hline 

\scriptsize 49 & \scriptsize Krishna \etal~\cite{beetle} & \scriptsize TSE & \scriptsize 2020 & \scriptsize Whence to Learn? Transferring Knowledge in Configurable [...] & \scriptsize X & \scriptsize X & \scriptsize X & \scriptsize X	\tabularnewline \hline 

\scriptsize 50 & \scriptsize Weber \etal~\cite{weber2021whitebox} & \scriptsize ICSE & \scriptsize 2021 & \scriptsize White-Box Performance-Influence Models: A Profiling [...] & \scriptsize X & \scriptsize X & \scriptsize  & \scriptsize 	\tabularnewline \hline 

\scriptsize 51 & \scriptsize Mühlbauer \etal~\cite{9285664} & \scriptsize ASE & \scriptsize 2020 & \scriptsize Identifying Software Performance Changes Across Variants [...] & \scriptsize X & \scriptsize X & \scriptsize  & \scriptsize 	\tabularnewline \hline 

\scriptsize 52 & \scriptsize Han \etal~\cite{han2020automated} & \scriptsize Report & \scriptsize 2020 & \scriptsize Automated Performance Tuning for Highly-Configurable [...] & \scriptsize X & \scriptsize X & \scriptsize  & \scriptsize 	\tabularnewline \hline 

\scriptsize 53 & \scriptsize Han \etal~\cite{10.1145/3427921.3450255} & \scriptsize ICPE & \scriptsize 2021 & \scriptsize ConfProf: White-Box Performance Profiling of Configuration [...] & \scriptsize X & \scriptsize  & \scriptsize X & \scriptsize 	\tabularnewline \hline 

\scriptsize 54 & \scriptsize Valov \etal~\cite{10.1145/3358960.3379127} & \scriptsize ICPE & \scriptsize 2020 & \scriptsize Transferring Pareto Frontiers across Heterogeneous Hardware [...] & \scriptsize X & \scriptsize  & \scriptsize  & \scriptsize X	\tabularnewline \hline 

\scriptsize 55 & \scriptsize Liu \etal~\cite{10.1145/3387902.3392633} & \scriptsize CF & \scriptsize 2020 & \scriptsize Deffe: a data-efficient framework for performance [...] & \scriptsize X & \scriptsize X & \scriptsize X & \scriptsize X	\tabularnewline \hline 

\scriptsize 56 & \scriptsize Fu \etal~\cite{265059} & \scriptsize NSDI & \scriptsize 2021 & \scriptsize On the Use of ML for Blackbox System Performance Prediction & \scriptsize X & \scriptsize X & \scriptsize X & \scriptsize \tabularnewline \hline 

\scriptsize 57 & \scriptsize Larsson \etal~\cite{9472818} & \scriptsize IFIP & \scriptsize 2021 & \scriptsize Source Selection in Transfer Learning for Improved Service [...] & \scriptsize X & \scriptsize X & \scriptsize X & \scriptsize X	\tabularnewline \hline 

\scriptsize 58 & \scriptsize Chen \etal~\cite{9401979} & \scriptsize ICSE & \scriptsize 2021 & \scriptsize Efficient Compiler Autotuning via Bayesian Optimization & \scriptsize X & \scriptsize X & \scriptsize X & \scriptsize 	\tabularnewline \hline 

\scriptsize 59 & \scriptsize Chen \etal~\cite{8787029} & \scriptsize SEAMS & \scriptsize 2019 & \scriptsize All Versus One: An Empirical Comparison on Retrained [...] & \scriptsize X & \scriptsize X & \scriptsize  & \scriptsize 	\tabularnewline \hline 

\scriptsize 60 & \scriptsize Ha \etal~\cite{8811988} & \scriptsize ICSE & \scriptsize 2019 & \scriptsize DeepPerf: Performance Prediction for Configurable Software [...] & \scriptsize X & \scriptsize  & \scriptsize  & \scriptsize 	\tabularnewline \hline 

\scriptsize 61 & \scriptsize Pei \etal~\cite{10.1145/3361566} & \scriptsize Report & \scriptsize 2019 & \scriptsize DeepXplore: automated white box testing of deep [...] & \scriptsize X & \scriptsize X & \scriptsize  & \scriptsize 	\tabularnewline \hline 

\scriptsize 62 & \scriptsize Ha \etal~\cite{8919029} & \scriptsize ICSME & \scriptsize 2019 & \scriptsize Performance-Influence Model for Highly Configurable [...] & \scriptsize X & \scriptsize  & \scriptsize  & \scriptsize 	\tabularnewline \hline 

\scriptsize 63 & \scriptsize Iorio \etal~\cite{8968941} & \scriptsize CloudCom & \scriptsize 2019 & \scriptsize Transfer Learning for Cross-Model Regression in Performance [...] & \scriptsize X & \scriptsize X & \scriptsize X & \scriptsize X \tabularnewline \hline 

\scriptsize 64 & \scriptsize Koc \etal~\cite{satune} & \scriptsize ASE & \scriptsize 2021 & \scriptsize SATune: A Study-Driven Auto-Tuning Approach for [...]	& \scriptsize X & \scriptsize X & \scriptsize X & \scriptsize X \tabularnewline \hline 

\scriptsize 65 & \scriptsize Ding \etal~\cite{intertransf} & \scriptsize ESEC/FSE & \scriptsize 2021 & \scriptsize Generalizable and Interpretable Learning for [...]	& \scriptsize X & \scriptsize X & \scriptsize X & \scriptsize X \tabularnewline \hline 

\multicolumn{5}{|c|}{\scriptsize Total} & \scriptsize \num{61}	& \scriptsize \num{39} & \scriptsize \num{29} & \scriptsize \num{15} \tabularnewline \hline
\end{tabular}
\end{table*}

In this section, we explore the significance of the input sensitivity problem in research. 
Do researchers know the issue of input sensitivity? 
How do they deal with inputs in their papers? 
%Do they ignore it? 
Is the interaction between software configurations and input sensitivity a well-known issue? 
%What motivates us is to estimate to what extent input sensitivity is a problem in the research community, but it is also the opportunity to promote innovative and original state-of-the-art solutions. 

\subsection{Experimental Protocol}
\label{sub:significance:protocol}

First, we aim at gathering research papers~\cite{10.1145/3368089.3409742} predicting the performance of configurable systems \ie with a performance model~\cite{guo2013}. 

\textbf{Gather research papers.} 
We focused on the publications of the last ten years.
To do so, we analyzed the papers published (strictly) after 2011 from the survey of Pereira \etal \cite{julianasurvey} - published in 2019.
We completed those papers with more recent papers (2019-2021), following the same procedure as in \cite{julianasurvey}. 
We have only kept research work that trained performance models.

\textbf{Search for input sensitivity.} 
We read each selected paper and answered four different questions:
Q-A. \pqone~ 
If not, the impact of input sensitivity in the existing research work would be relatively low. 
The idea of this research question is to estimate the proportion of the performance models that could be affected by input sensitivity.
Q-B. \pqtwo~ 
If not, it would suggest that the performance model only captures a partial truth, and might not generalize for other inputs fed to the software system. 
Q-C. \pqthree~ 
This question aims to state whether researchers are aware of the input sensitivity issue, and estimate the proportion of the papers that mention it as a potential threat to validity. 
Q-D. \pqfour~ 
Finally, we check whether the paper proposes a solution managing input sensitivity \ie if the proposed approach could be adapted to our problem and predict a near-optimal configuration for any input. 
The results were obtained by one author and validated by all other co-authors.

\subsection{How do Research Papers Address Input Sensitivity?}
\label{sub:significance:results}

\Cref{tab:papers} lists the \nbpapers research papers we identified following this protocol, as well as their individual answers to Q-A$\to$Q-D. A checked cell indicates that the answer to the corresponding question (column) for the corresponding paper (line) is \textit{yes}. Since answering Q-B, Q-C or Q-D only makes sense if Q-A is checked, we grayed and did not consider Q-B, Q-C and Q-D if the answer of Q-A is \textit{no}. We also provide full references and detailed justifications in the companion repository\footnote{List of papers at \scriptsize{\bluerl{https://github.com/llesoil/input_sensitivity/tree/master/results/RQS/RQ6/}}}. 
We now comment the average results:

%\begin{enumerate}[label=$Q-{\Alph*}.$ :,topsep=0pt,itemsep=-1ex,partopsep=1ex,parsep=1ex]

\textbf{Q-A.} \pqone~ Of the \nbpapers papers, \num{60} (\pc{94}) consider at least one configurable system processing inputs. This large proportion gives credits to input sensitivity and its potential impact on research work.

\textbf{Q-B.} \pqtwo~ \pc{63} of the research work answering \textit{yes} to Q-A include different inputs in their protocol. But what about the other \pc{37}? It is understandable not to consider several inputs because of the cost of measurements. However, if we reproduce all experiments of \Cref{tab:papers} using other input data, will we draw the same conclusions for each paper? Based on the results of $RQ_{1}$ $\to$ $RQ_{3}$, we encourage researchers to consider at least a set of inputs in their protocol (see \Cref{sec:discuss}).

\textbf{Q-C.} \pqthree~ Only half (\pc{47}) of the papers mention the issue of input sensitivity, mostly without naming it or using a domain-specific keyword \eg workload variation \cite{DBLP:conf/wosp/ValovPGFC17}. For the other half, we cannot guarantee with certainty that input sensitivity concerns all papers. But we shed light on this issue: ignoring input sensitivity can prevent the generalization of performance models across inputs. This is especially true for the \pc{37} of papers answering \textit{no} to Q-B \ie considering one input per system: only \pc{14} of these research works mention it. 

\textbf{Q-D.} \pqfour~ We identified \nbsolutionpapers papers~\cite{10.1145/3035918.3064029,ding2015,10.1145/3194133.3194138, satune, jamshidi2017b, DBLP:conf/wosp/ValovPGFC17,jamshidi2018,beetle,10.1145/3358960.3379127,9472818,8968941,10.1145/2684200.2684314,lillack2013improved,10.1145/3387902.3392633,intertransf} proposing contributions that may help in better managing the input sensitivity problem.
However, most of them have not been designed to operate over actual inputs, but rather changes of computing environments~\cite{jamshidi2017b,DBLP:conf/wosp/ValovPGFC17,jamshidi2018,beetle,10.1145/3358960.3379127,9472818,8968941}. Other works~\cite{10.1145/3035918.3064029,ding2015,10.1145/3194133.3194138, satune,10.1145/3387902.3392633} only apply it to a specific domain (database~\cite{10.1145/3035918.3064029}, compilation~\cite{ding2015}, cloud computing~\cite{10.1145/3194133.3194138, 10.1145/3387902.3392633,intertransf} or programs analysis~\cite{satune}) with open questions about applicability and effectiveness in other areas. 
We plan to confront these techniques on our dataset for multiple systems.  
\vspace*{-0.2cm}

\begin{tcolorbox}[boxsep=-2pt]
\textbf{Conclusion.}
While half of the research articles mention input sensitivity, few actually address it, and most often on a single system and domain. 
Input sensitivity can affect multiple research works and questions their practical relevance for a field deployment. 
%Some state-of-the-art candidate solutions are not systematically applicable and their cost and accuracy must be assessed. 
% On the bright side, we could adapt and apply state-of-the-art techniques for this problem. 
% It is yet an open challenge to tune a configurable system for its input data. 
\end{tcolorbox}

\vspace*{-0.2cm}

\section{Implications of our study}
\label{sec:discuss}

% Our study has several implications regarding 
% This section discusses the implications of our work.
In this section, we first summarize results of our study and then discuss their impacts on several research directions.

\begin{table}[h]
\renewcommand{\arraystretch}{0.9}
\setlength{\tabcolsep}{1.7pt}
\centering
% \vspace*{-0.2cm}
\caption{Summary of the input sensitivity on our dataset}
%Due to space constraints, we arbitrarly select few performance properties.
\label{tab:recap}
% \vspace*{-0.2cm}
% \hspace*{-0.1cm}
\begin{tabular}{|c|c|c|c|c|c|c|c|c|c|c|}
\hline
\multirow{3}{*}{\scriptsize System}
&  \multirow{3}{*}{\scriptsize Perf.}
&  \scriptsize Correlation
&  \scriptsize Most infl. opt.
&  \scriptsize Impact (\%)
&  \scriptsize \textbf{IS}
\tabularnewline 

&  
&  \scriptsizeb{ [$C_{min}$, $C_{max}$]}
&  \scriptsizeb{ Effect [min, max]}
&  \scriptsizeb{ $100*(Q_{2}-1)$}
&  \scriptsize \textbf{Score}
\tabularnewline

&  
&  
&  
&  
&  \scriptsize (0$\to$1)
\tabularnewline 

&  
&  \scriptsize ($RQ_{1}$)
&  \scriptsize ($RQ_{2}$)
&  \scriptsize ($RQ_{3}$)
&  \scriptsize \textbf{($RQ_{4}$)}
\tabularnewline \hline
\multirow{3}{*}{\scriptsize \soft{gcc}}
& \scriptsize \performance{ctime}
&  \scriptsize [$0.72$, $0.97$]
&  \scriptsize [$-0.26$, $-0.18$]
&  \scriptsize $13$
&  \scriptsize \textbf{0.32}
\tabularnewline

& \scriptsize \performance{exec}
&  \scriptsize [$-0.69$, $1$]
&  \scriptsize [$-0.21$, $0.64$]
&  \scriptsize $27$
&  \scriptsize \cellcolor[HTML]{808080}{\textbf{0.92}}
\tabularnewline 

& \scriptsize \performance{size}
&  \scriptsize [$0.48$, $1$]
&  \scriptsize [$-0.03$, $0$]
&  \scriptsize $7$
&  \scriptsize \textbf{0.27}
\tabularnewline \hline
\scriptsize \soft{image}
& \scriptsize \performance{time}
&  \scriptsize [$-0.24$, $1$]
&  \scriptsize [$-0.18$, $0.95$]
&  \scriptsize $4$
&  \scriptsize \textbf{0.39}
\tabularnewline \hline
\multirow{2}{*}{\scriptsize \soft{lingeling}}
& \scriptsize \performance{\# conf}
&  \scriptsize [$-0.9$, $0.92$]
&  \scriptsize [$-0.79$, $0.91$]
&  \scriptsize $15$
&  \scriptsize \cellcolor[HTML]{e8e8e8}{\textbf{0.75}}
\tabularnewline 
& \scriptsize \performance{\# reduc}
&  \scriptsize [$-0.99$, $1$]
&  \scriptsize [$-0.79$, $0.91$]
&  \scriptsize $10$
&  \scriptsize \cellcolor[HTML]{e8e8e8}{\textbf{0.7}}
\tabularnewline \hline
\scriptsize \soft{nodejs}
& \scriptsize \performance{ops}
&  \scriptsize [$-0.87$, $0.95$]
&  \scriptsize [$-1<$, $0.93$]
&  \scriptsize $17$
&  \scriptsize \cellcolor[HTML]{e8e8e8}{\textbf{0.79}}
\tabularnewline \hline
\multirow{2}{*}{\scriptsize  \soft{poppler}}
& \scriptsize \performance{size}
&  \scriptsize [$-1$, $1$]
&  \scriptsize [$-1<$, $>1$]
&  \scriptsize $7$
&  \scriptsize \cellcolor[HTML]{e8e8e8}{\textbf{0.64}}
\tabularnewline 

& \scriptsize \performance{time}
&  \scriptsize [$-0.94$, $1$]
&  \scriptsize [$-0.93$, $>1$]
&  \scriptsize $37$
&  \scriptsize \cellcolor[HTML]{808080}{\textbf{0.98}}
\tabularnewline \hline
\multirow{2}{*}{\scriptsize \soft{SQLite}}
& \scriptsize \performance{q1}
&  \scriptsize [$-0.78$, $0.87$]
&  \scriptsize [$-0.69$, $0.58$]
&  \scriptsize $2$
&  \scriptsize \textbf{0.45}
\tabularnewline

& \scriptsize \performance{q15}
&  \scriptsize [$-0.3$, $0.94$]
&  \scriptsize [$-0.59$, $0.6$]
&  \scriptsize $3$
&  \scriptsize \textbf{0.37}
\tabularnewline \hline
\multirow{5}{*}{\scriptsize \soft{x264}}
& \scriptsize \performance{bitrate}
&  \scriptsize [$-0.69$, $1$] 
&  \scriptsize [$-0.55$, $0.28$]
&  \scriptsize $21$
&  \scriptsize \cellcolor[HTML]{808080}{\textbf{0.84}}
\tabularnewline 

& \scriptsize \performance{cpu} 
&  \scriptsize [$-0.31$, $1$]
&  \scriptsize [$-0.1$, $0.84$]
&  \scriptsize $7$
&  \scriptsize \textbf{0.47}
\tabularnewline 

& \scriptsize \performance{fps}
&  \scriptsize [$0.01$, $1$]
&  \scriptsize [$-0.62$, $0.23$]
&  \scriptsize $6$
&  \scriptsize \textbf{0.37}
\tabularnewline 

& \scriptsize \performance{size}
&  \scriptsize [$-0.69$, $1$]
&  \scriptsize [$-0.58$, $0.28$]
&  \scriptsize $21$
&  \scriptsize \cellcolor[HTML]{808080}{\textbf{0.84}}
\tabularnewline 

& \scriptsize \performance{time}
&  \scriptsize [$0.02$, $1$]
&  \scriptsize [$-0.17$, $0.45$]
&  \scriptsize $7$
&  \scriptsize \textbf{0.39}
\tabularnewline \hline
\multirow{2}{*}{\scriptsize \soft{xz}}
& \scriptsize \performance{size}
&  \scriptsize [$0.14$, $1$]
&  \scriptsize [$-0.02$, $0.94$]
&  \scriptsize $0$
&  \scriptsize \textbf{0.22}
\tabularnewline
& \scriptsize \performance{time}
&  \scriptsize [$-0.03$, $0.97$]
&  \scriptsize [$-0.98$, $-0.15$]
&  \scriptsize $6$
&  \scriptsize \textbf{0.37}
\tabularnewline \hline
\end{tabular}
\vspace*{-0.2cm}
\end{table}

\subsection{Synthesis and interpretation of results}
\label{subsec:synth}

To support the discussions, we rely on a table that summarizes the major results of $RQ_{1}$, $RQ_{2}$ and $RQ_{3}$ by providing different indicators per software system and per performance property. 
Specifically, \Cref{tab:recap} reports the standard deviation of Spearman correlations (as in $RQ_{1}$), the minimal and maximal effects of the most influential option (as in $RQ_{2}$), the average relative difference of performance due to inputs (as in $RQ_{3}$).  
\begin{comment}
This table quantifies the input sensitivity of software systems along different perspectives:
\begin{enumerate}[label=$\bullet$,topsep=0pt,itemsep=-1ex,partopsep=1ex,parsep=1ex, leftmargin=0.2cm]
    \item \textbf{generalization of configuration knowledge:} indicators of $RQ_{1}$ and $RQ_{2}$ are mainly here to determine whether a performance model over configurations generalizes well to an arbitrary input. For instance, systems like \soft{xz} or \soft{imagemagick} have strong correlations (greater than 0.8 on average with weak standard deviations less than 0.2) for their \performance{size} and \performance{time}.  
    In contrast, a system like \soft{poppler} is more sensitive to input with weak correlation (less than 0.2; standard deviation greater than 0.4). 
    \item \textbf{performance improvements of configurations:} indicators of $RQ_{3}$ reported in \Cref{tab:recap} but also in \Cref{tab:ratios}, page~\pageref{tab:ratios} are here to show how much performance can be gained when the input is taken into account. 
    For instance, inputs of \soft{poppler} can drastically speed up performance. In contrast, inputs of \soft{xz} for \performance{size} have no effect \ie performance remain stable whatever the input. 
\end{enumerate}
\end{comment}
Out of the results of \Cref{tab:recap}, we can make further observations. First, input sensitivity is specific to both a configurable system and a performance property. For instance, the sensitivity of \soft{x264} configurations differs depending on whether \performance{bitrate} or \performance{cpu} are considered. 
Second, there are configurable systems for which inputs threaten the generalization of configuration knowledge, but the performance ratios remain affordable (\eg \soft{SQLite} for \performance{q1}).
Intuitively, one needs a way to assess the level of input sensitivity per system and per performance property. 
We propose a metric that aggregates both indicators of $RQ_{1}$, $RQ_{2}$ and $RQ_{3}$. 
% We first need to choose a threshold $\alpha$, representing the maximal proportion of variability due to inputs we can tolerate. 
% For instance, if we consider that \pc{5} of lost performance is acceptable, $\alpha$ will be fixed at \num{0.05}. 
 We define the score of $I$nput $S$ensitivity as follows:
\begin{center}
%\smallskip
\noindent$IS = \frac{1}{4}*|C_{max} - C_{min}| + \frac{1}{2\alpha}*\min(Q_{2}-1,\alpha)$
%\smallskip
\end{center}

%\begin{itemize}[topsep=0pt,itemsep=-1ex,partopsep=1ex,parsep=1ex, leftmargin=0.2cm]\item 
where $C_{min}$ and $C_{max}$ are the minimal and maximal Spearman correlations
%\item 
$Q_{2}$ is the median of the performance ratio distribution, and $\alpha$ a threshold representing the maximal proportion of variability due to inputs we can tolerate.
%$Q_{1}$ and $Q_{3}$ are \resp the first and third quartiles of the performance ratios.
%\end{itemize}
The first part of the formula quantifies (in $[0, 0.5]$, as $|C_{max} - C_{min}|$ is in $[0, 2]$) how the input sensitivity changes the configuration knowledge ($RQ_{1}$ and $RQ_{2}$). 
For instance, a textbook case of software system with no input sensitivity would have only performance correlations of \num{1}, leading to a first part equal to $\frac{1}{4}*|1-1| = 0$. 
But if the correlations are completely opposite between different inputs, this first part would be equal to $\frac{1}{4}*|1-(-1)| = \frac{2}{4} = 0.5$.
The second part quantifies (in $[0, 0.5]$) the impact of input sensitivity ($RQ_{3}$) in the actual performance. 
For instance, a software system with no impact of input sensitivity would have only performance ratios equals to \num{1}, leading to $Q_{2} = 1$ and $\frac{1}{2\alpha}*\min(Q_{2}-1,\alpha) = 0$. 
Conversely, for high performance ratios, $Q_{2}-1 >> \alpha$, $\min(Q_{2}-1,\alpha) = \alpha$ and $\frac{1}{2\alpha}*\min(Q_{2}-1,\alpha) = \frac{\alpha}{2\alpha} = 0.5$
%Both are positive and majored by \num{0.5}, 
$IS$ thus varies between \num{0} (no input sensitivity) and \num{1} (high input sensitivity). 
We compute $IS$ for each couple of systems and performance properties of our dataset, with $\alpha$ fixed at \pc{25}(see \Cref{tab:recap}).
Empirical evidences show that $IS$ values are robust and trustworthy when using the measurements of \num{15} inputs or more\footnote{See \bluerl{https://github.com/llesoil/input_sensitivity/tree/master/results/RQS/RQ5/RQ5-evolution.ipynb}}. 

$IS$ scores are reported in \Cref{tab:recap} as follows: systems and performance properties with scores higher than \num{0.5} as input-sensitive (lightgray), and those with $IS$ greater than \num{0.8} as highly input-sensitive (gray). 
$IS$ scores highlight the input sensitive cases \eg \num{0.98} for the \performance{time} of \soft{poppler}, \num{0.84} for the \performance{bitrate} and the \performance{size} of \soft{x264}.
Systems like \soft{xz} or \soft{imagemagick} exhibit low $IS$ scores that reflect their low sensitivity to inputs.
As a small validation, we also compute the $IS$ of \soft{x264} for input videos used in \cite{10.1145/3358960.3379137}. 
We retrieve scores of \num{0.31} and \num{0.66} for the \performance{time} and the \performance{size} of \soft{x264}\footnote{See \bluerl{https://github.com/llesoil/input_sensitivity/tree/master/results/RQS/RQ5/RQ5-other_ref.ipynb}}. 
% Overall, most of the configurable systems are input-sensitive. 
% We now discuss the implications of our results on several research directions. 
% the input sensitivity score $IS$, as defined in the protocol of $RQ_{4}$. 
% Based on our work, we propose to consider systems and performance properties with scores higher than \num{0.5} as input-sensitive (lightgray), and those with $IS$ greater than \num{0.8} as highly input-sensitive (gray). 

%\fixme{IS score per group x264} 
% -> bonne idée, mais peut-être qu'on a déjà beaucoup de contenu 
% TODO Luc

\subsection{Implications, insights, open challenges}

Our study has several implications for different tasks related to the performance of software system configurations. 
For each task, we systematically discuss the key insights and open problems brought by our results and not addressed in the state of the art.

\textbf{Tuning configurable systems.} Numerous works aim to find optimal configurations of a configurable system. 
\newline
\emph{Key insights.} Our empirical results show that the best configuration can be differently ranked (see $RQ_{1}$) depending on an input. 
The tuning cannot be reused as such, but should be redone or adapted whenever a system processes a new input. 
Another key result is that it is worth taking input into account when tuning: relatively high performance gains can be obtained (see $RQ_{3}$). 
\newline
\emph{Open challenges.}
The main challenge is thus to deliver algorithms and practical tools capable of tuning the performance of a system, whatever the input. 
A related issue is to minimize the cost of tuning. For instance, tuning from scratch -- each time a new input is fed to a software system -- seems impractical since too costly.
Approaches that reuse configuration knowledge through \eg prioritized sampling or transfer learning can be helpful here, but should be carefully assessed \wrt costs and actual performance improvements.  

\textbf{Performance prediction of configurable systems.} Numerous works aim to predict the performance of an arbitrary configuration. 
\newline
\emph{Key insights.} Looking at indicators of $RQ_{1}$ and $RQ_{2}$, inputs can threaten the generalization of configuration knowledge. 
That is, a performance prediction model trained out of one input can be highly inaccurate for many other inputs.  
\newline
\emph{Open challenges.} The ability to transfer configuration knowledge across inputs is a critical issue. 
Transfer learning techniques have been explored, but mostly for hardware or version changes~\cite{DBLP:conf/wosp/ValovPGFC17,9555247} and not for inputs' changes. 
Such techniques require measuring several configurations each time an input is targeted. 
It also requires training performance models that can be reused.  
Owing to the huge space of possible inputs, this computational cost can be a barrier if systematically applied. 
A possible direction for reducing measurements' cost is to group together similar inputs.

%This issue has been partly considered but in specific domains (SAT solvers \cite{xu2008, Falkner2015SpySMACAC}, compilation \cite{inputs_compilation,ding2015}, video encoding \cite{Maxiaguine2004}, data compression \cite{8820983}, \etc). It is unclear whether the solutions are cost-effective and generalizable to all domains and software systems. Overall, more research is needed to systematically support input-aware tuning of software systems. 

%\textbf{Impacts for practitioners.}
%Interpretability and u
\textbf{Understanding of configurable systems.} Understanding the effects of options and their interactions is hard for developers and users yet crucial for maintaining, debugging or configuring a software system. 
Some works (\eg \cite{weber2021whitebox}) have proposed to build performance models that are interpretable and capable of communicating the influence of individual options on performance.  % (possibly back to the code).and interactions
\newline
\emph{Key insights.}
Our empirical results show that performance models, options and their interactions are sensitive to inputs (see indicators of $RQ_{2}$). 
To concretely illustrate this, we present a minimal example using SPLConqueror~\cite{10.1007/s11219-011-9152-9} a tool to synthesize interpretable models. We trained two performance models predicting the encoding \performance{sizes} of two different input videos fed to \soft{x264}. Unfortunately, the two related models do not share any common (interaction of) option\footnote{See the performance models for \textcolor{blue}{\href{https://github.com/llesoil/input_sensitivity/tree/master/results/splconqueror/model_Animation_1080P-3d67.txt}{the first}} and \textcolor{blue}{\href{https://github.com/llesoil/input_sensitivity/tree/master/results/splconqueror/modelAnimation_720P-0acc.txt}{the second}} input videos.}. Let us be clear: the fault lies not with SPLConqueror, but with the fact that a model simply does not generalize to any input. 
\newline
\emph{Open challenges.} Hence, a first open issue is to communicate when and how options interact with input data. The properties of the input can be exploited, but they must be understandable to developers and users. 
Another challenge is to identify a minimal set of representative inputs (see $RQ_{4}$) in such a way interpretable performance models can be learnt out of observations of configurable systems.

\textbf{Effectiveness of sampling and learning strategies}. % Impacts for researchers.} 
Measuring a few configurations (a sample) to learn and predict the performance of any configurations has been subject to intensive research. 
The problem is to sample a small and representative set of configurations and inputs that leads to a good accuracy. % of performance prediction models. 
% Input sensitivity further exacerbates the problem.
\newline
\emph{Key insights.}
 A key observation of $RQ_{2}$ is that the importance of options can vary across inputs. 
Therefore, sampling strategies that prioritize or neglect some options may miss important observations if the specifics of inputs are not considered. 
% Practically, a sampling strategy may well be highly accurate for an input and still inaccurate for others. 
We thus warn researchers that the effectiveness of sampling strategies for a given configurable system can be biased by the inputs and the performance property used.
\newline
\emph{Open challenges.} % We warn researchers that the effectiveness of sampling strategies for a given configurable system can be biased by the inputs and the performance property used.
% In Section~\ref{sec:significance}, we show that most of the studies neglect either inputs or configurations, which is perfectly understandable owing to the investments required. 
% A notable exception is 
 Pereira \etal~\cite{10.1145/3358960.3379137} showed that some sampling strategies are more or less effective depending on the 19 videos and 2 performance properties of \soft{x264}. 
% Our study considers much more videos for \soft{x264} (1397), subject systems (beyond \soft{x264}) and performance properties. 
Kaltenecker \etal~\cite{DBLP:journals/software/KalteneckerGSA20} empirically showed that there is no one-size-fits-all solution when choosing a sampling strategy together with a learning technique. 
 We suspect that input sensitivity further exacerbates the phenomenon.
% The scientific community should be extremely careful with this input sensitivity issue when assessing the effectiveness of sampling strategies. 
 Using our dataset, we are seeing two opportunities for researchers: (1) assessing state-of-the-art sampling strategies; (2) designing input-aware sampling strategies \ie cost-effective for any input. 
% In view of the results of our study, new problems deserve to be tackled with associated challenges. We detail some of them hereafter. 
% , a prediction or optimisation algorithm, or a transfer technique 

%Another consequence of our results is that practitioners should measure the impacts of inputs on configurations and react accordingly, which may take different forms: 
%(1) default configurations that neglect inputs' specifics should be reconsidered; 
%(2) the documentation should be as clear as possible to inform or even guide users; 
%(3) for testing software performance (\eg ensuring non-regression in a continuous integration system), a set of representative inputs should be used.

\textbf{Testing and benchmarking configurable systems.} With limited budget, developers continuously test the performance of configurable systems for ensuring non-regression.  
%  As the testing budget is limited, there is a need to choose which configurations to measure and over which input data 
\newline
\emph{Key insights.} Testing software configurations on a single, fixed input can hide several interesting insights related to software properties (\eg performance bugs). 
From this perspective, indicators of $RQ_{2}$ about influences of options should be analyzed and controlled. 
Similarly, performance drop (see $RQ_{3}$) should be handled. 
\newline
\emph{Open challenges.} 
% \textit{(2) Selecting representative inputs.}
To reduce the cost of measurements, the ideal would be to select a set of input data, both representative of the usage of the system and cheap to measure.
We believe our work (see $RQ_{4})$ can be helpful here.
On the \soft{x264} case study, for the \performance{bitrate}, we isolate four encoding groups of input videos - see \Cref{tab:input_group} in $RQ_{2}$. 
Within a group, the videos share common properties, and \soft{x264} processes them in the same way \ie same performance distributions ($RQ_{1}$), same options' effects ($RQ_{2}$) and a negligible impact of input sensitivity ($RQ_{3}$). 
Automating this grouping could drastically reduce the cost of testing. An approach applicable to any kind of input and configurable software is yet to be defined and assessed. 
 
 \textbf{Detecting input sensitivity.} Practitioners and scientists should have the means to determine whether a software under study is input-sensitive \wrt the performance property of interest.
% \textbf{Empirical knowledge.} We conduct a large empirical study over several systems, software configurations, performance properties, and input data.
%  \newline
\newline
\emph{Key insights.} % The sensitivity to input has been overlooked 
We propose several indicators (as part of $RQ_{1}$, $RQ_{2}$, and $RQ_{3}$) as well as $IS$ a simple, aggregated score to quantify the level of input sensitivity. 
Such metrics can be leveraged to take inform decisions as part of the tasks previously discussed. 
% For instance, in case input sensitivity is low (\eg for \soft{xz} and its two properties) a limited number of inputs can be considered. % . 
%If the input sensitivity is negligible (see $RQ_{3}$), we can use a single model to predict the performance of the software system. 
% Unfortunately, the general observation is that input sensitivity can be more severe and high: multiple inputs should be considered. 
%  multiple inputs are needed.
\newline
\emph{Open challenges.} Detecting input sensitivity has a computational cost. Selecting the right subset of configurations and input data is thus a key issue. 
Our empirical experiments (see Section~\ref{subsec:synth}) suggest that a limited percentage of inputs (around \pc{10}) can be used to quantify sensitivity. 
Other indicators and metrics can also be proposed to quantify sensitivity to inputs.  
 Our study is the first to providence evidence of input sensitivity. We also share data with \num{1976025} measurements that can be analyzed and reused to consolidate configuration knowledge. 
However, further empirical knowledge is more than welcome to understand the significance of input sensitivity on other software systems and performance properties.  

% \textbf{Recommendations for researchers and practitioners.} Given the state of the art and the open problems to be addressed, there is no complete solution that can be systematically employed. 
% However, we can give two recommendations: \textit{(1) Detecting input sensitivity.}

% Knowledge about the software and domain should be leveraged to select and group representative inputs.  

% and kill the input variability?

%\textbf{Other variability factors.} \fixme{S'il nous reste de la place, mais un peu HS; parler d'une combinaison input avec hardware, os etc en gros deep var centrée sur les inputs.} 

\section{Threats to Validity} \label{sec:threats}

This section discusses the threats to validity related to our protocol.

\textbf{Construct validity.} 
Due to resource constraints, we did not include all the options of the configurable systems in the experimental protocol. 
We may have forgotten configuration options that matter when predicting the performance of our configurable systems.
However, we consider features that impact the performance properties according to the documentation, which is sufficient to show the existence of the input sensitivity issue. 
The use of random sampling also represents a threat, in the sense that the measured configurations could not be representative of a real-world usage of the software systems. To mitigate this threat, we toke care of selecting documented and informed options, typically part of custom configurations and profiles, that are supposed to have an effect of performance. We mainly relied on documentation and guides associated to the projects.
The validity of the conclusions can depend on the choice of systems under test. In the context of \cite{lesoil2022transferring}, we conducted an additional experiment to ensure the robustness of our results for \soft{x265}, an alternative software to \soft{x264}. Results\footnote{See at  \url{https://github.com/llesoil/input_sensitivity/blob/master/results/others/x264_x265/x264_x265.ipynb}} show that the performance distributions are different from \soft{x264} to \soft{x265} (except for \performance{size}) but the input sensitivity problem holds for \soft{x265} when it is observed for \soft{x264}.

\textbf{Internal Validity.}
%%%%%%%%%%%%%%%%%%%%%%%%%%%% MEASUREMENT PROCESS
% measurement bias
First, our results can be subject to measurement bias. 
We alleviated this threat by making sure only our experiment was running on the server we used to measure the performance of software systems. 
It has several benefits: we can guarantee we use similar hardware (both in terms of CPU and disk) for all measurements; we can control the workload of each machine (basically we force the machine to be used only by us); we can avoid networking and I/O issues by placing inputs on local folders. 
But it could also represent a threat: our experiments may depend on the hardware and operating system. 
To mitigate this, we conducted an additional experiment on \soft{x264} over a subset of inputs to show the robustness of results whatever the hardware platforms\footnote{See  the companion repository at \url{https://github.com/llesoil/input_sensitivity/blob/master/results/others/x264_hardware/x264_hardware.ipynb}}. 
% docker
The measurement process is launched via Docker containers. 
% (see the "Docker" column in \Cref{tab:xp_systems})
If this aims at making this work reproducible, this can also alter the results of our experiment. 
% repeat the measures
Because of the amount of resources needed to compute all the measures, we did not repeat the process of \Cref{fig:measurement_process} several times per system. 
We consider that the large number of inputs under test overcomes this threat. 
Moreover, related work (\eg \cite{10.1145/3358960.3379137} for \soft{x264}) has shown that inputs lead to stable performance measurements across different launches of the same configuration. 
%\fixme{Additionally, we conducted experiments studying the performance consistency over a  subset of inputs, and report on the results in the companion repository\footnote{\fixme{TODO si le temps}}.}
Finally, the measurement process can also suffer from a lack of inputs. 
To limit this problem, we took relevant dataset of inputs produced and widely used in their field. 
%%%%%%%%%%%%%%%%%%%%%%%%%%%% RQS
%For $RQ_{1}$-$RQ_{3}$, executing our code with another python environment may lead to slightly different conclusions. 
For $RQ_{3}$, we consider oracles when predicting the best configurations for both scenarios, thus neglecting the imprecision of performance models: these results might change on a real-world case. 
In \Cref{sec:significance}, our results are subject to the selection of research papers: since we use and reproduce \cite{julianasurvey}, we face the same threats to validity.

\textbf{External Validity.}
A threat to external validity is related to the used case studies and the discussion of the results.
Because we rely on specific systems and interesting performance properties, the results may be subject to these systems and properties. 
To reduce this bias, we selected multiple configurable systems, used for different purposes in different domains. 

\section{Related Work} \label{sec:related_work}

In this section, we discuss other related work (see also Section~\ref{sec:significance}).

%\textbf{Machine learning and configurable systems.}
%Machine learning techniques have been widely considered in the literature to learn software configuration spaces \cite{julianasurvey,quinton2020evolution,beetle, jamshidi2017b, jamshidi2018, DBLP:conf/wosp/ValovPGFC17, nair2017, 10.1145/3106237.3106273,FSE2017menzies,flash_find_config, velez2021whitebox}.
%Numerous works have proposed to model performance of software configurations, with several use-cases in mind for developers and users of software systems: the maintenance and understanding of configuration options and their interactions \cite{SGKA:ESECFSE15}, the selection of an optimal configuration \cite{10.1145/3106237.3106273,FSE2017menzies,flash_find_config}, the automated specialization of configurable systems \cite{temple:hal-01467299,IEEEcontextTemple}. 

\textbf{Workload Performance Analysis.}
On the one hand some work have been addressing the performance analysis of software systems \cite{blockchain_platform, Coppa2014, image_sensitivity, Goldsmith2007, leitner2016, DBLP:conf/ssbse/SinhaCC20} depending on different input data (also called workloads or benchmarks), but all of them only considered a rather limited set of configurations. 
On the other hand, as already discussed in \Cref{sec:significance}, works and studies on configurable systems usually neglect input data (\eg using a unique video for measuring the configurations of a video encoder). 
In this paper, we combined both dimensions by performing an in-depth, controlled study of several configurable systems to make it vary in the large, both in terms of configurations and inputs. 
In contrast to research papers considering multiple factors of the executing environment in the wild \cite{jamshidi2018, DBLP:conf/wosp/ValovPGFC17}, we concentrated on inputs and software configurations only, which allowed us to draw reliable conclusions regarding the specific impact of inputs on software variability.

\textbf{Performance Prediction.}
% One word on current solutions
Research work have shown that machine learning could predict the performance of configurations \cite{guo2013,guo2015,DBLP:conf/kbse/ZhangGBC15,DBLP:conf/splc/ValovGC15}.
These works measure the performance of a configuration sample under specific settings to then build a model capable of predicting the performance of any other configuration, \ie a performance model.
Numerous works have proposed to model performance of software configurations, with several use-cases in mind for developers and users of software systems: the maintenance and understanding of configuration options and their interactions \cite{SGKA:ESECFSE15}, the selection of an optimal configuration \cite{10.1145/3106237.3106273,FSE2017menzies,flash_find_config}, the automated specialization of configurable systems \cite{temple:hal-01467299,IEEEcontextTemple}.
Input sensitivity complicates their task; since inputs affect software performance, it is yet a challenge to train reusable performance prediction models \ie that we could apply on multiple inputs.

\textbf{Input-aware tuning.} 
The input sensitivity issue has been partly considered in some specific domains (SAT solvers \cite{xu2008, Falkner2015SpySMACAC}, compilation \cite{inputs_compilation,ding2015}, video encoding \cite{Maxiaguine2004}, data compression \cite{8820983}, \etc). 
%In each field, experts try to optimize the performance of a software system according to a given workload. 
%It is unclear whether these ad-hoc solutions are cost-effective and generalizable to all domains and software configurations. For example, is it always possible and effective to extract input properties for all kinds of inputs? 
% Overall, more research is needed to systematically support input-aware tuning of configurable software.
It is unclear whether these ad hoc solutions are cost-effective. 
As future work, we plan to systematically assess domain-specific techniques as well as generic, domain-agnostic approach (\eg transfer learning) using our dataset.
Furthermore, the existence of a general solution applicable to all domains and software configurations is an open question. For example, is it always possible and effective to extract input properties for all kinds of inputs?

\textbf{Input Data and other Variability Factors.} 
Most of the studies support learning models restrictive to specific static settings (\eg inputs, hardware, and version) such that a new prediction model has to be learned from scratch once the environment change \cite{julianasurvey}. 
%The study of Valov \etal \cite{DBLP:conf/wosp/ValovPGFC17} suggests that changing the hardware has reasonable impacts since linear functions are highly accurate when reusing prediction models.
Jamshidi \etal \cite{jamshidi2017b} conducted an empirical study on four configurable systems (including \soft{SQLite} and \soft{x264}), varying software configurations and environmental conditions, such as hardware, input, and software versions. 
But without isolating the individual effect of input data on software configurations, it is challenging to understand the existing interplay between the inputs and any other variability factor \cite{10.1145/3442391.3442402} \eg the hardware.

\section{Conclusion} \label{sec:conclusion}

We conducted a large study over the inputs fed to \nbsystems configurable systems that shows the significance of the input sensitivity problem on performance properties. 
We deliver one main message: \textbf{inputs interact with configuration options in non-monotonous ways, thus making it difficult to (automatically) configure a system}. 

There are also some opportunities when tackling the input sensitivity problem. 
We have shown it is possible to select a representative set of inputs and thus to greatly reduce the cost of benchmarking software \eg \num{1397} $\to$ \num{8} inputs for \soft{x264} despite high sensitivity.  

Our analysis of the literature showed that input sensitivity has been either overlooked or partially addressed. % on specific domains. 
We have pointed out several open problems to consider related to tuning, prediction, understanding, and testing of configurable systems. 
In light of the results of our study, we encourage researchers to confront existing methods and explore future ideas with our dataset. %address the input sensitivity problem. 
 
% Our analysis of the literature showed that input sensitivity has been partially addressed, and several open problems are to be considered. 

As future work, it is an open challenge to solve the issue of input sensitivity when predicting, tuning, understanding, or testing configurable systems. In particular, a direct follow-up work aim at adapting the current practice of performance models to overcome input sensitivity and train models robusts to the change of input data.

\begin{comment}
\vspace*{-0.2cm}
\begin{tcolorbox}[boxsep=-2pt]
\begin{center}
\textbf{TAKEAWAY MESSAGES}
\end{center}
\begin{itemize}[topsep=0pt,itemsep=-1ex,partopsep=1ex,parsep=1ex, leftmargin=0.1cm]
    \item Input sensitivity relates to a majority of software systems;
    \item In average, ignoring it leads to a performance drop of \pc{38}. We cannot completely ignore input sensitivity;
    \item We propose a score detecting input sensitivity;
    \item Selecting a representative set of inputs greatly reduces the cost of benchmarking software \eg \num{1397} $\to$ \num{8} inputs for \soft{x264};
    \item Most of research works do not address the problem of input sensitivity - or on a specific domain. 
\end{itemize}
\end{tcolorbox}
\end{comment}

%\vspace*{-0.3cm}

\begin{comment}
input{sections/01-introduction}
input{sections/02-problem}
input{sections/03-protocol-bis}
input{sections/04-results-bis}
input{sections/05-significance}
input{sections/06-discussion}
input{sections/07-threats}
%input{sections/08-related}
input{sections/09-conclusion}
\end{comment}

\nolinenumbers

%\balance
\bibliographystyle{spmpsci}
\bibliography{bib/inputs}
%\bibliography{inputs.bbl}

\end{document}